**Keldysh Institute of Applied Mathematics**
**of Russian Academy of Sciences**

**V.D. Lakhno**

# Translational-invariant bipolarons and superconductivity

**V.D. Lakhno**

**Translational-invariant bipolarons and superconductivity**

A translation-invariant (TI) bipolaron theory of superconductivity based, like Bardeen–Cooper–Schrieffer theory, on Fröhlich Hamiltonian is presented. Here the role of Cooper pairs belongs to TI bipolarons which are pairs of spatially delocalized electrons whose correlation length of a coupled state is small. The presence of Fermi surface leads to stabilization of such states in its vicinity and a possibility of their Bose-Einstein condensation (BEC).

The theory provides a natural explanation of the existence of a pseudogap phase preceding the superconductivity and enables one to estimate the temperature of a transition $T^*$ from a normal state to a pseudogap one.

It is shown that the temperature of BEC of TI bipolarons determines the temperature of a superconducting transition $T_c$ which depends not on the bipolaron effective mass but on the ordinary mass of a band electron. This removes restrictions on the upper limit of $T_c$ for a strong electron-phonon interaction. A natural explanation is provided for the angular dependence of the superconducting gap which is determined by the angular dependence of the phonon spectrum.

It is demonstrated that a lot of experiments on thermodynamic and transport characteristics, Josephson tunneling and angle-resolved photoemission spectroscopy (ARPES) of high-temperature superconductors does not contradict the concept of a TI bipolaron mechanism of superconductivity in these materials. Possible ways of enhancing $T_c$ and producing new room-temperature superconductors are discussed on the basis of the theory suggested

## Contents

## 1. Introduction

The theory of superconductivity for ordinary metals is one of the finest and long-established branches of condensed matter physics which involves macroscopic and microscopic theories and derivation of macroscopic equations from a microscopic description [1]. In this regard the theory at its core was presented in its finished form and its further development should imply detalization and consideration of special cases.

The situation changed after the discovery of high-temperature superconductivity (HTSC) [2]. Suddenly, the correlation length in oxide ceramics turned out to be several orders of magnitude less than in ordinary metal superconductors and the width of the gap – much larger than the superconducting transition temperature [3]. The current state of the theory and experiment is given in books and reviews [4]-[15].

Presently the main problem is to develop a microscopic theory capable of explaining experimental facts which cannot be explained by the standard Bardeen–Cooper–Schrieffer theory (BCS) [16].

While modern versions of a microscopic description of HTSC are many – phonon, plasmon, spin, exciton, etc. – the central point of a microscopic theory is the effect of electron coupling (Cooper effect). Such "bosonization" of electrons further lies in the core of the description of their superconducting condensate.

The phenomenon of pairing in a broad sense is the formation of bielectron states and in a narrow sense, if the description is based on a phonon mechanism – the formation of bipolaron states. For a long time this concept has been in conflict with a great correlation length or the size of Cooper pairs in the BCS theory. The same reason hindered the treatment of superconductivity as a boson condensate (see footnote at page 1177 in [16]). In no small measure this incomprehension was caused by a standard idea of bipolarons as very compact formations.

The first indication of the fallacy of this viewpoint was obtained in work [17] where an analogy between the BCS and Bose-Einstein condensation (BEC) was demonstrated while studying the properties of a high-density exciton gas. The results of [17] enabled one to develop the idea of a crossover, i.e. passing on from the BCS theory which corresponds to the limit of weak electron-phonon interaction (EPI) to the BEC theory which corresponds to the limit of strong EPI [18]-[24]. It was believed that additional evidence in favor of this way is Eliashberg strong coupling theory [25]. According to [26], in the limit of infinitely strong EPI this theory leads to the regime of local pairs, though greatly different from the BEC regime [27].

However, attempts to develop a crossover theory between BCS and BEC faced insurmountable difficulties. For example, it was suggested to develop a theory, with the use of a T-matrix approach, where the T-matrix of the initial fermion system would transform into the T-matrix of the boson system as the EPI enhances [28]-[33]. However, this approach turned out to fail even in the case of heavily diluted systems. Actually the point is that even in the limit when a system consists of only two fermions one cannot construct a one-boson state of them. In the EPI theory this problem is known as the bipolaron one.

One reason why the crossover theory failed is as follows. Like the bipolaron theory, the BCS theory is based on the Fröhlich Hamiltonian. For this Hamiltonian, an important theorem of the analyticity of the polaron and bipolaron energy with respect to EPI constant is proved [34]. However, in the BCS theory an important assumption is made – a real matrix element is replaced by a model quantity which is matrix element a truncated from the top and from the bottom of phonon momenta. This procedure is, by no means, fair. As it is shown in [35], in the bipolaron theory this leads to side effects – existence of a local energy level separated by a gap from the quasicontinuous spectrum (Cooper effect). This solution is isolated and non-analytical with respect to the coupling constant. In the BCS theory just this solution forms the basis for the development of the superconductivity theory.

As a result, the theory developed and its analytical continuation – Eliashberg theory – distort the reality and, in particular, make it impossible to construct a theory on the basis of the BEC. Replacement of a real matrix element by its model analog enables one to perform analytical calculations completely. In particular, a replacement of a real interaction by a local one in the BCS enabled one to derive the phenomenological Ginzburg-Landau model which is also a local model [36]. Actually, the power of this approach can hardly be overestimated since it enabled one to get a lot of statements consistent with the experiment.

Another more important reason why the crossover theory failed is that vacuum in the polaron (bipolaron) theory with spontaneously broken symmetry differs from the vacuum in the translation invariant (TI) polaron (TI bipolaron) theory in the case of strong interaction which makes impossible for Eliashberg theory to pass on to the strong coupling TI bipolaron theory (Section 2).

In this review we present a superconductivity (SC) theory based on the electron-phonon interaction. There the BCS corresponds to the limit of weak EPI (Section 3), and the case of strong EPI corresponds to a TI bipolaron superconductivity theory where the SC phase corresponds to a TI bipolaron Bose condensate.

The relevance of a review on a bipolaron mechanism of superconductivity is caused by the following facts: 1) Most reviews of bipolaron SC are devoted to small-radius polarons (SRP) [37], while in the past time, after the theory of SC on the basis of SRP had been criticized [38]-[41] interest has shifted to large-radius polarons 2) Most of papers published in the past decades were devoted to magnet-fluctuation mechanisms of SC while more recent experiments where record $T_c$ (under high pressure) were obtained, were performed on hydrogen sulfides and lantanium hydrides where magnetic interactions are lacking but there is a strong electron-phonon interaction 3) Crucial evidence in favor of a bipolaron mechanism is provided by recent experiments [42] which demonstrate the existence of pairs at temperatures higher than $T_c$. 4) Important evidence for the bipolaron mechanism of SC is experiments (Božović et al. [43]) where the number of paired states in high-temperature superconductors (HTSC) was demonstrated to be far less that the total number of current carriers.

In this review we outline the main ideas of translation invariant (TI) polarons and bipolarons in polar crystals. In review [44] they are presented in greater detail. As in the Bardeen–Cooper–Schrieffer theory, the description of the TI bipolaron gas is based on the electron-phonon interaction and Fröhlich Hamiltonian. As distinct from the BCS theory where the correlation length greatly exceeds the mean distance between the pairs, in this review we deal with the opposite case when the correlation length is far less than the distance between the pairs.

The thermodynamical characteristics of a three-dimensional TI bipolaron Bose condensate are reviewed. The critical transition temperature, energy, heat capacity, transition heat of a TI bipolaron gas are discussed.

The influence of an external magnetic field on the thermodynamic characteristics of a TI bipolaron gas is considered. We compare with the experiment such characteristics as: maximum value of the magnetic field intensity for which a TI bipolaron condensate can exist, London penetration depth and temperature dependence of these quantities. The results obtained are used to explain experiments on high-temperature superconductors.

Special attention is given to the fact that, according to TI bipolaron theory of HTSC, different types of experiments measure different quantities as a SC gap. We show that in tunnel experiments the bipolaron energy is found while in angular resolved photo electron spectroscopy (ARPES) phonon frequency for which EPI is maximum is measured. According to the TI bipolaron theory of SC, a natural explanation is provided for such phenomena as the occurrence of kinks in the spectral measurements of a gap, angular dependence of a gap, the availability of a pseudogap, etc.

## 2. Polaron and fundamental problems of non-relativistic quantum field theory

The polaron theory is based on the Fröhlich Hamiltonian which describes the interaction of an electron with phonon field:

$$H = \frac{\hat{\boldsymbol{p}}^2}{2m} + \sum_k \hbar\omega(k)\, a_k^+ a_k + \sum_k V_k \left[a_k e^{ikr} + a_k e^{-ikr}\right], \tag{1}$$

where $\vec{r}$ is the radius-vector of an electron, and $\hat{\vec{p}}$ is its momentum; m is the electron effective mass; $a_k^+$, $a_k$ are operators of the birth and annihilation of the field quanta with energy ħω(k) , $V_k$ is the matrix element of an interaction between an electron and a phonon field.

In the condensed matter physics the polaron theory is a broad field which involves the description of electron properties of ionic crystals [45]-[49], polar superconductors [50]-[51], conducting polymers [52]-[54], biopolymers [55]-[56], high-temperature superconductors [57]-[59], magnetic semiconductors [60] and other important objects of condensed matter.

The reason of such popularity of the polaron model is its universality. Fundamentally, all the physical phenomena are described relying on the quantum-field formulation. In non-relativistic physics its simplest realization is based on the use of Fröhlich Hamiltonian (1).

Various expressions for $V_k$ and $\omega(k)$ in the case of ionic crystals, piezoelectrics, superconductors, nuclear matter, degenerated semiconductor plasma are given in [61]. In the past years Hamiltonian (1) has been used for the description of impurity atoms placed into the Bose-Einstein condensate of ultracold atoms [62], electrons in low-dimensional systems [59], [63] etc.

Rather a simple form of Hamiltonian (1) has encouraged researchers to find an exact solution of the polaron problem. In the stationary state an exact solution would give a spectrum of Hamiltonian (1) and, as a consequence, a solution of a wide range of condensed matter physics problems. However, the problem turned out to be much more complicated than it seemed to be. To solve it, use was made of various methods and techniques of the quantum field theory such as the Green function method, diagram technique, path integral method, renormalization group method, quantum Monte Carlo method, diagram Monte Carlo method, etc. Various variational approaches, the most efficient of which turned out to be Feynman's path integral method, enabled researchers to find an approximate dependence of the polaron ground state energy over the whole range of variation of the electron-phonon interaction constant α.

The above approaches, however, failed to determine the spectrum of Hamiltonian (1) even in the weak coupling limit [64].

In the limit of strong coupling, in order to investigate the properties of Hamiltonian (1), starting with pioneering works by Pekar [45], use was made of the canonical transformation:

$$a_k \rightarrow a_k - V_k \rho_k^* / \hbar\omega(k), \tag{2}$$

where $\rho_k^*$ is the Fourier-component of the charge distribution density. Transformation (2) singles out the classical component (second term in the right-hand side of (2)) from the quantum field which, by assumption, should make the main contribution into the strong coupling limit. Starting with the work by Lieb [65] (see also [66] and references therein) a lot of papers dealt with the proof that the functional of Pekar total energy for the polaron ground state yielded by (2) is asymptotically exact in the strong coupling limit. In other words, it was argued that the choice of a variation wave function of the ground state in the form:

$$|\Psi\rangle = \varphi(r) \exp \sum_k V_k \frac{\rho^*}{\hbar\omega(k)} (a_k - a_k^+) |0\rangle , \tag{3}$$

where φ(r) is the electron wave function, which, in the case of the total energy of the polaron ground state E= <Ψ|H|Ψ>, leads to Pekar functional of the strong coupling :

$$E = \frac{\hbar^2}{2m} \int |\nabla\Psi|^2 \, d^3r - \sum_k \frac{V_k^2}{\hbar\omega(k)} \rho_k^* \rho_k , \tag{4}$$

yields rather an exact solution in the strong coupling limit.

In this case the spectrum of the polaron excited states was considered only for a resting polaron $P$=0, where $P$ is the polaron momentum. Variation of (4) over $\Psi^*$ leads to a nonlinear Schroedinger equation for the wave function Ψ, which has the form of Hartree equation. Numerical integration of this equation was performed in [67] and some polaron excited states and relevant renormalized phonon modes were found (see, for example, review [61]).

Hence, most of the papers on the polaron theory in the strong coupling limit realize the method of quantizing in the vicinity of a classical solution which is now widely used in the non-perturbative quantum field theory [68].

Fundamentally, this method seemed unsatisfactory even at the early stages of the development of the polaron theory. Indeed, if in the strong coupling limit the polarization field can be considered to be classical and non-zero, it becomes unclear how this macroscopic state can be held by a single electron. It is possible only in the case when the field is equal to zero except for a small region where the electron is localized forming a self-consistent state with the field. In this case the initial translational symmetry turns out to be broken: the polarization potential well can spontaneously form, equally likely, in any region of space. All the attempts to construct a translationally invariant theory on the basis of this physical picture yielded the same results that the initial semiclassical strong-coupling theory developed by Pekar [45].

The situation changed radically after the publication of papers [69]-[71] where a fundamentally different mechanism of an electron motion in a polar crystal was considered. According to [69]-[71], when moving along a crystal, an electron not only displaces equilibrium states of atoms, but also alters the profile of their potential energy in the crystal which is equivalent to the formation of their squeezed oscillatory states [73,74]. For a TI polaron, average (i.e. classical) displacements of atoms from their equilibrium positions, as distinct from a Pekar polaron, are equal to zero. Accordingly, polarization of the crystal is equal to zero too, since a TI polaron is spatially localized. However, the mean values of phonon occupation numbers in a polaron crystal are not equal to zero. This paradox is resolved by the fact that the non-zero mean number of phonons is caused by the availability of squeezed (i.e. non-classical) states excited by an electron.

Squeezing of phonon states induced by the electron motion along a crystal leads to a new type of a bound state of the electron-phonon system described by a unified wave function which presents a new type of ansatz and cannot be presented as a factorized ansatz formed by the electron and phonon parts individually.

The theory of squeezed states was first used in the polaron theory in Tulub work [75]. In view of a non-optimal choice of the variation wave function in [75], the results obtained for the ground state energy actually reproduce those derived by Pekar. This significantly delayed their use in the polaron theory. For this reason intensive development of the squeezed state theory took place much later – after the paper by Glauber [76], who drew attention to their important role in the understanding of the principle of uncertainty and the principle of superposition in quantum mechanics.

Presently squeezed states have widespread application: in optics they are used to suppress self-noise of light, in computing technics – for the development of optical computers and communication lines, in precision measurements – in interference antennae of gravitation waves, etc. (see, for example, books and reviews [77] - [79]).

In the polaron theory the squeezed state method, after the pioneering work by Tulub [75], was used in works [80]- [84] for a discrete model of a Holstein polaron [85] and in works [86-93] for a Fröhlich polaron. In works [80] - [84] some very important results were obtained: first, the polaron ground state energy of squeezed states turned out to be lower than that in all the papers on a Holstein polaron where an ordinary vacuum is used, second, the effective mass of a Holstein polaron calculated for squeezed states appeared to be much less than that obtained by Holstein [85].

This is not the case with works [86] - [93] where the squeezed state theory was applied to Fröhlich Hamiltonian. Despite the fact that there a considerable enhancing of polaron effects was observed when squeezed states were used, in general, the results did not differ from those obtained by Pekar [45]. As mentioned above, breakthrough results were obtained in [69] - [71] where for Pekar-Fröhlich Hamiltonian, it was shown that the energy of a polaron ground state and the energy of a bipolaron for squeezed states is lower than that in the Pekar theory.

The most important application of the polaron and bipolaron theory is superconductivity. Apparently, development of a superconductivity theory is the most difficult problem of the condensed matter physics since it requires a solution of a multiparticle problem. This problem was solved by Bardeen, Cooper and Schrieffer in the limit of weak interaction on the basis of Fröhlich Hamiltonian (1) [16] Section 3. Its solution enabled one to explain some properties of ordinary superconductors.

The discovery of high-temperature superconductivity (HTSC) showed that the BCS theory probably cannot be applied to them since the electron-phonon interaction in HTSC materials cannot be considered to be weak. Presently, to describe this case researchers resort to the use of Eliashberg theory [25], [94]-[95], since it was developed for the case of a strong electron-phonon interaction and in the weak coupling limit it coincides with the BCS theory. However the use of Eliashberg theory in the case of HTSC had limited success. This fact gave rise to a number of theories which were based not on the Hamiltonian of electron-phonon interaction (1) but on other types of interaction different from EPI. These works, eventually, faced the same problems that Eliashberg theory did.

In Eliashberg theory a small parameter is a ratio $\omega/E_F$, where ω is the phonon frequency, $E_F$ is the Fermi energy. If $\omega/E_F \to 0$, then the electron-phonon interaction constant $\alpha \sim \omega^{-1/2} \to \infty$. The perturbation theory with respect to this parameter is developed for ordinary vacuum phonon functions | 0 >, which are taken as a zero approximation. But for $\alpha \to \infty$, the proper choice of the zero approximation will be the function $\Lambda_0 | 0 >$,

$$\hat{\Lambda}_0 = c \exp\left\{\frac{1}{2}\sum_{k,k'} a_k^+ A_{kk'} a_{k'}^+\right\}, \tag{5}$$

where $\Lambda_0$ is the operator of a squeezed state [44], [75], [96], [97]. Hence, the Eliashberg theory developed for ordinary vacuum will give different results than the theory developed for squeezed vacuum does. Obviously, in the limit of weak coupling when $\alpha \to 0$, the results of both the theories will coincide. However, as α increases, boson vacuum determined by the vacuum function | 0 > will be more and more unstable and for a certain critical value $\alpha_c$, a new boson vacuum determined by the function $\hat{\Lambda}_0 | 0 >$ emerges. It will be a lower energy state. The inapplicability of the Migdal theorem was probably first mentioned by Alexandrov [37], who, in relation to the superconductivity theory based on small-radius polarons, pointed out that vacuum chosen on the basis of Migdal theorem “knows nothing” about another vacuum which is a polaron narrowing of the conductivity band and formation of a SRP in a new vacuum of squeezed states [80]-[84]. For this reason Eliashberg theory is inadequate for explanation of HTSC.

The foundation of superfluidity was laid in papers by London and Tisza [98], [99], who were the first to relate the fundamental phenomenon of Bose-Einstein condensation to phenomenon of superfluidity. The idea to treat superconductivity as superfluidity of electron liquid was suggested by Landau in [100]. There a spectrum of elementary excitations of a superfluid liquid was introduced for the first time which received the name of a roton spectrum and enabled one to construct a statistical mechanics of a superfluid state. Landau could not transfer the ideas of his work on superfluidity to superconductivity because of a difference in statistics: Fermi statistics for electrons in metals and Bose statistics for helium atoms in liquid helium II. The work by Bogolyubov on superconductivity [101] which related the phenomenon of Bose condensate to superfluidity could have accelerated the construction of the superconductivity theory on the basis of Bose condensate, however at that time there was not an example of Bose gas of charged bosons which is necessary for superconductivity.

Further sequence of events is well known: in 1950 Ginzburg and Landau developed a phenomenological theory of superconductivity [102] in which a microscopic mechanism of superconductivity was not discussed since its possible nature was unclear.

Finally, in the work by Bardeen, Cooper and Schrieffer [16] a microscopic mechanism of superconductivity was found. This was the mechanism of Cooper pairing of electrons. Cooper pairs, being bosons, supposedly could have played the role of particles from which Bose gas consists and, thus, have combined the theories of superconductivity and superfluidity. However, that did not happen. The answer was given in the BCS theory *per se* – the size of Cooper pairs in metals turned out to be so huge that in each pair there was about of a million of other pairs. For this reason an analogy between a Bose-Einstein condensate and superconductivity was discarded in [16]. Interest in it emerged only in 1986 when Müller and Bednorz discovered high-temperature superconductivity.

To be fair it should be emphasized that a possibility of the formation of such a singular quantum state as Bose condensate was predicted by Einstein on the basis of generalization of Bose statistics to the case of a finite mass of a Bose particle. Until the publication of the BCS work there was not an example of a charged boson with a finite mass in the condensed matter physics. The first example of a possible existence of such quasiparticles was a Cooper pair which enabled BCS to construct a theory of superconductivity. A Cooper pair, as was mentioned above, being overlapped with others could not be a true quasiparticle. For the same reason, both in the BCS and in Bogolyubov theory [103], there is only a single-electron spectrum of Fermi-type excitations. Hence in the framework of BCS, as it was stated by its authors, a theory of Bose condensate cannot be constructed.

In 70 – 80 of the last century a small-radius bipolaron (SRBP) was considered as a quasiparticle possessing the properties of a charged boson, having a mass, and capable of forming a Bose-condensate in narrow-bandgap crystals [104].

For a long time works on superconductivity based on the idea of Bose condensate of SRBP were developed by Alexandrov and his colleagues [37], [58], [40], [105].

In view of a large mass of SRP and SRBP the temperature of the SC transition determined by the temperature of BEC formation should be low. This fact was pointed out in papers [38],[39],[41], which criticized the SC theory based on SRP.

After the discovery of high-temperature superconductivity some other approaches were developed the most popular of which was Anderson resonating valence bond theory (RVB) and *t-J* model [106]-[107].

Notwithstanding a strong attraction of these models from the viewpoint of theory, for example, a possibility to describe both conducting and magnetic properties of crystals on the basis of one simple Hamiltonian, they turned out to be ineffective for explaining HTSC. In particular the fact of a possible existence of a SC phase in these models did not receive a reliable proof.

In view of the fact that recent experiments [42]-[43] suggest a phonon nature of the superconductivity mechanism in HTSC with a record $T_c$, further presentation is based on EPI. Being general, the theoretical approaches considered can be applied to other types of interaction different from EPI.

## 3. Weak EPI. BCS theory

In BCS a multielectron problem is solved on the assumption that electrons interact only with a phonon field and do not interact with one another. Hence only an ensemble of independent electrons in a phonon field is considered. Such a picture of BCS is substantiated by a Fermi-liquid model of a metal which implies that instead of strongly interacting electrons we can consider non-interacting quasiparticles, i.e. an ideal Fermi gas in a phonon field. In this case, the one-electron Fröhlich Hamiltonian (1) can be written in the form suitable for the description of any number of electrons:

$$H = \sum_{p,s} \varepsilon_p c^+_{p,s} c_{p,s} + \sum_q \omega(q) a^+_q a_q + \sum_{\substack{p,q,s \\ p'-p=q}} V_q c^+_{p,s} c_{p's} a^+_q + H.c., \qquad (6)$$

$$\varepsilon_p = p^2/2m - E_F \, , \omega(q) = s_0 q \; ,$$

where $c^+_{p,s}$, $c_{p,s}$ are operators of the birth and annihilation of electrons with momentum p and spin s, $s_0$is the sound velocity. In ( 6 ) the energy of electron states is reckoned from the Fermi level $E_F$.

In the case of metals for which the BCS is used:

$$V_q = G(\omega(q)/2V)^{1/2} \; ,$$

$G$ is the interaction constant. For a weak EPI, using the perturbation theory we can exclude phonon operators $a^+_q$, $a_q$ and present (6) as the Hamiltonian:

$$H = \sum_{p,s} \varepsilon_p c^+_{p,s} c_{ps} + \sum_{p,p',k,s,s'} V^2_k \frac{\hbar\omega(k)}{(\varepsilon_{p+k} - \varepsilon_k)^2 - \hbar^2\omega^2(k)} c^+_{p+k,s} c^+_{p'-k,s'} c_{p',s'} c_{p,s} \qquad (7)$$

In the BCS theory an important approximation is made: it is believed that the main contribution into the interaction is made only by the processes occurring in the energy range $|\varepsilon_p - \varepsilon_{p'}| < \hbar\omega_D$, in the vicinity of the Fermi level where $\omega_D$ is the Debye frequency of a phonon. In this energy range the coefficient preceding the electron operators in the interaction term is replaced by the constant $g$.

The BCS theory is based on the choice of a probe function in the form of a superposition of Cooper pairs with $p = -p', s = -s'$. Hence, in the BCS instead of (7) consideration is given to the Hamiltonian:

$$H = \sum_{p,s} \varepsilon_p c^+_{p,s} c_{ps} - g \sum_{p,k} c^+_{p+k,\uparrow} c^+_{-p-k,\downarrow} c_{-p\downarrow} c_{p\uparrow} =$$

$$\sum_{p,s} \varepsilon_p c^+_{p,s} c_{p,s} - g \sum_p c^+_{p\uparrow} c^+_{-p\downarrow} \sum_{p'} c_{-p'\downarrow} c_{p'\uparrow}. \qquad (8)$$

Hamiltonian (8) can be diagonalized via the canonical transformation:

$$c_{p\uparrow} = u_p \xi_{p\uparrow} + v_p \xi^+_{-p,\downarrow}, \;\; c_{-p\downarrow} = u_p \xi_{-p\downarrow} - v_p \xi^+_{p\uparrow} \, ,$$

$$c^{+}_{p\uparrow} = u_p \xi^{+}_{p\uparrow} + v_p \xi_{-p\downarrow}\,, \; c^{+}_{-p\downarrow} = u_p \xi^{+}_{-p,\downarrow} - v_p \xi_{p,\uparrow}\,. \tag{9}$$

As a result, Hamiltonian (8) is written as:

$$H = E_0 + \sum_p E_p \left(\xi^{+}_{p\uparrow}\xi_{p\uparrow} + \xi^{+}_{-p\downarrow}\xi_{-p\downarrow}\right), \tag{10}$$

$$E_p = \sqrt{\left(\frac{p^2}{2m} - E_F\right)^2 + \Delta^2}\,, \qquad \Delta = 2g \sum_p{}' u_p\, v_p, \qquad u_p^2 = 1 - v_p^2 = (1 + \mathcal{E}_p/E_p)/2,$$

where the prime in the expression for Δ means that summation is performed over the states lying in a thin layer of the Fermi surface where interaction is non-zero, $|v_p|^2$ provides a probability that the state (p↑, - p↓) is occupied, and $|u_p|^2$ is the probability that it is free.

The results obtained correspond only to the case of T=0. In particular, the energy of the ground state of the system under consideration reckoned from the energy of the system in the normal state (i.e. with Δ=0) is equal to:

$$E = <\Psi|H|\Psi> = -\frac{1}{2}N(0)\Delta^2, \qquad \Delta = 2\hbar\omega_D \exp\left(-\frac{1}{N(0)g}\right),$$

$$|\Psi> = \prod_{p_1 \dots p_N}(u_p + v_p c^{+}_{p\uparrow} c^{+}_{-p\downarrow})|0>\,, \tag{11}$$

where N(0) is the electron density at the Fermi level in a normal phase, N is the number of electrons.

Hence, formation of paired states leads to a decrease of the system energy by the value of $N(0)\Delta^2/2$ and emergence of superconductivity. It follows from (10) that the density of elementary excitations is $\rho(E_p) \to \infty$ for $E_p \to \Delta$ . In the TI bipolaron theory of SC this corresponds to the formation of a Bose condensate of paired electrons with an infinite state density for the energy equal to the bipolaron energy which is separated by a gap from the continuous excitation spectrum.

The problem of the number of paired electrons, i.e. Cooper pairs in the BCS theory is treated differently by different authors. For example, it is often argued (see for example [108]) that electrons are paired only in the narrow layer of the Fermi surface so that their number $N_S$ is equal to $N_S = \frac{\Delta}{E_F}N$ . For $\Delta/E_F \simeq 10^{-4}$ only a small portion of electrons are paired.

The BCS theory, however, gives an unambiguous answer: for T=0, $N_S = \frac{N}{2}$ (which straightforwardly follows from the expression for the wave function (11)), i.e. all the electrons are in the paired state.

To resolve this contradiction let us consider the contribution of $w_p$ into the total energy of a superconductor which is made by a pair in the state (p↑, - p↓):

$$w_p = \mathcal{E}_p - E_p \tag{12}$$

It follows from (12) that $w_{p_F} = -\Delta$ . In the normal state (Δ=0) such a pair would contribute the energy $w^N_{p_F} = 0$, that is $\delta w_{p_F} = w^s_{p_F} - w^N_{p_F} = -\Delta$. Accordingly, at the bottom of the conductivity band, i.e. for p=0 expression (12) is written as: $w^s_0 = -2E_F - \Delta^2/2E_F$. It follows that pairs occurring far below the Fermi surface outside the layer of width Δ in the BCS approximation make a very small contribution into the SC energy which is approximately $\delta w_0/\delta w_{P_F} = \Delta/2E_F \sim 10^{-4}$ of the contribution made by the pairs in the layer Δ. Hence, though all the electrons in the BCS are paired their contribution, depending on the energy of the pair, is different. Only in the thin layer of the Fermi surface it is nonzero. This fact just resolves the above contradiction: though all the electrons are paired, the energy is contributed only by a small number of the pairs: $N_s = N\Delta/E_F$, which is called a number of pairs in a SC.

Therefore, in order not to make a mistake in calculating some or other characteristic of a SC, when a solution is not obvious, one should take account of all the paired states of the electrons.

For example, when calculating London depth of the magnetic field penetration into a SC in the BCS theory, one should take account of all the electron paired states. At the same time, when calculating the critical magnetic field for which a SC deteriorates, it is sufficient to estimate them only in the layer Δ.

It should be noted that in recent experiments by Božović [43] it was shown that only a small portion of electrons make a contribution into the London penetration depth in HTSC. This means that the BCS theory is inapplicable to them and the interaction cannot be considered to be weak. This problem will be considered in Section 11.

## 4. Pekar-Fröhlich Hamiltonian. Canonical transformations.

Before we pass on to presentation of the SC theory in the limit of strong EPI, let us outline the results of the TI bipolaron theory.

In describing bipolarons, according to [69]-[71],[109], we will proceed from Pekar-Fröhlich Hamiltonian in a magnetic field:

$$H = \frac{1}{2m}\left(\hat{\boldsymbol{p}}_1 - \frac{e}{c}\boldsymbol{A}_1\right)^2 + \frac{1}{2m}\left(\hat{\boldsymbol{p}}_2 - \frac{e}{c}\boldsymbol{A}_2\right)^2 + \sum_k \hbar\omega_k^0 a_k^+ a_k + \quad (13)$$

$$\sum_k (V_k e^{ikr_1} a_k + V_k e^{ikr_2} a_k + H.c.) + U(|\boldsymbol{r}_1 - \boldsymbol{r}_2|),$$

$$U(|\boldsymbol{r}_1 - \boldsymbol{r}_2|) = \frac{e^2}{\epsilon_\infty |\boldsymbol{r}_1 - \boldsymbol{r}_2|},$$

where $\hat{p}_1, \vec{r}_1, \hat{p}_2, \vec{r}_2$ are momenta and coordinates of the first and second electrons, $\vec{A}_1 = \vec{A}(\vec{r}_1)$, $\vec{A}_2 = \vec{A}(\vec{r}_2)$ are vector-potentials of the external magnetic field at the points where the first and second electrons occur; $U$ describes Coulomb repulsion between the electrons. We write Hamiltonian (13) in general form. In the case of HTSC which are ionic crystals $V_k$ is a function of the wave vector $k$ which corresponds to the interactions between the electrons and optical phonons:

$$V_k = \frac{e}{k}\sqrt{\frac{2\pi\hbar\omega_0}{\tilde{\epsilon}V}} = \frac{\hbar\omega_0}{ku^{1/2}}\left(\frac{4\pi\alpha}{V}\right)^{1/2}, \qquad u = \left(\frac{2m\omega_0}{\hbar}\right)^{1/2}, \qquad \alpha = \frac{1}{2}\frac{e^2 u}{\hbar\omega_0\tilde{\epsilon}}, \quad (14)$$

$$\tilde{\epsilon}^{-1} = \epsilon_\infty^{-1} - \epsilon_0^{-1}, \quad \omega_k^0 = \omega_0,$$

where e is the electron charge; $\epsilon_\infty$ and $\epsilon_0$ are high-frequency and static dielectric permittivities; α is a constant of the electron-phonon interaction; V is the system volume, $\omega_0$ is a frequency of an optical phonon.

The axis *z* is chosen along the magnetic field induction $\vec{B}$ and symmetrical gauge is used:

$$\boldsymbol{A}_j = \frac{1}{2}\boldsymbol{B} \times \boldsymbol{r}_j$$

for *j* = 1,2. For the bipolaron singlet state considered below, the contribution of the spin term is equal to zero.

In the system of the mass center Hamiltonian (13) takes the form:

$$H = \frac{1}{2M_e}(\widehat{\boldsymbol{p}}_R - \frac{2e}{c}\boldsymbol{A}_R)^2 + \frac{1}{2\mu_e}(\widehat{\boldsymbol{p}}_r - \frac{e}{2c}\boldsymbol{A}_r)^2 + \sum_k \hbar\omega_k^0 a_k^+ a_k + \sum_k 2V_k \cos\frac{\boldsymbol{kr}}{2}\left(a_k e^{ikR} + H.c.\right) + U(|\boldsymbol{r}|), \quad (15)$$

$$\boldsymbol{R} = \frac{\boldsymbol{r}_1 + \boldsymbol{r}_2}{2}, \quad \boldsymbol{r} = \boldsymbol{r}_1 - \boldsymbol{r}_2, \quad M_e = 2m, \ \mu_e = m/2,$$

$$\boldsymbol{A}_r = \frac{1}{2}B(-y, x, 0), \quad \boldsymbol{A}_R = \frac{1}{2}B(-Y, X, 0),$$

$$\widehat{\boldsymbol{p}}_R = \widehat{\boldsymbol{p}}_1 + \widehat{\boldsymbol{p}}_2 = -i\hbar\nabla_{\boldsymbol{R}}, \quad \widehat{\boldsymbol{p}}_r = \frac{\widehat{\boldsymbol{p}}_1 - \widehat{\boldsymbol{p}}_2}{2} = -i\hbar\nabla_{\boldsymbol{r}},$$

where *x; y* and *X; Y* are components of the vectors $\vec{r}, \vec{R}$ accordingly.
Let us transform Hamiltonian *H* by Heisenberg transformation [110], [111]:

$$S_1 = \exp i\left(\boldsymbol{G} - \sum_k \boldsymbol{k}\, a_k^+ a_k\right)\boldsymbol{R}, \quad (16)$$

$$\boldsymbol{G} = \widehat{\boldsymbol{P}}_R + \frac{2e}{c}\boldsymbol{A}_R, \ \ \widehat{\boldsymbol{P}}_R = \widehat{\boldsymbol{p}}_R + \sum_k \hbar\boldsymbol{k} a_k^+ a_k, \quad (17)$$

where $\vec{G}$ commutates with the Hamiltonian, thereby being a constant, i.e. *c*-number, $\hat{\vec{P}}_R$ is the total momentum in the absence of the magnetic field.
Action of $S_1$on the field operator yields:

$$S_1^{-1} a_k S_1 = a_k e^{-ikR}, \quad S_1^{-1} a_k^+ S_1 = a_k^+ e^{ikR}. \quad (18)$$

Accordingly, the transformed Hamiltonian $\widetilde{H} = S_1^{-1} H S_1$ takes on the form:

$$\widetilde{H} = \frac{1}{2M_e}(\boldsymbol{G} - \sum_k \hbar\boldsymbol{k} a_k^+ a_k - \frac{2e}{c}\boldsymbol{A}_R)^2 + \frac{1}{2\mu_e}(\widehat{\boldsymbol{p}}_r - \frac{e}{2c}\boldsymbol{A}_r)^2 + \sum_k \hbar\omega_k^0 a_k^0 a_k + \sum_k 2V_k \cos\frac{\boldsymbol{kr}}{2}(a_k + a_k^+) + U(|\boldsymbol{r}|). \quad (19)$$

In what follows we will assume:

$$\boldsymbol{G} = 0. \quad (20)$$

The physical meaning of condition (20) is that the total momentum in a sample volume is equal to zero, i.e. a current is lacking. This requirement follows from the Meissner effect according to which the current in a SC volume should be equal to zero. We use this fact in section 8 in determining the London penetration depth λ.
Let us seek a solution of a stationary Schroedinger equation corresponding to Hamiltonian (19) in the form:

$$\Psi_H(r,R,\{a_k\}) = \phi(R)\Psi_{H=0}(r,R,\{a_k\}), \tag{21}$$

$$\phi(R) = \exp\left(-\mathrm{i}\frac{2e}{\hbar c}\int_0^{\boldsymbol{R}} \boldsymbol{A}_{R'}\, d\boldsymbol{R}'\right),$$
$$\Psi_{H=0}(r,R,\{a_k\}) = \psi(r)\Theta(R,\{a_k\}),$$

where $\Psi_{H=0}(r,R,\{a_k\})$ is the bipolaron wave function in the absence of a magnetic field. The explicit form of the functions $\psi$ $(r)$ and $\Theta(R,\{a_k\})$ is given in [44],[69]-[71].

Averaging of $\tilde{H}$ with respect to the wave functions $\phi(R)$ and $\Psi(r)$ yields:

$$\bar{\tilde{H}} = \frac{1}{2M_e}(\sum_k \hbar\boldsymbol{k}a_k^+ a_k)^2 + \sum_k \hbar\tilde{\omega}_k\, a_k^+ a_k + \tag{22}$$
$$\sum_k \bar{V}_k(a_k + a_k^+) + \bar{T} + \bar{U} + \bar{\Pi},$$

where:

$$\bar{T} = \frac{1}{2\mu_e}\langle\psi\left|\left(\hat{\boldsymbol{p}}_r - \frac{e}{2c}\boldsymbol{A}_r\right)^2\right|\psi\rangle, \qquad \bar{U} = \langle\psi|U(r)|\psi\rangle, \tag{23}$$

$$\bar{\Pi} = \frac{2e^2}{M_e c^2}\langle\phi|A_R^2|\phi\rangle$$

$$\hbar\tilde{\omega}_k = \hbar\omega_k^0 + \frac{2\hbar e}{M_e c}\langle\phi|\boldsymbol{k}\boldsymbol{A}_R|\phi\rangle.$$

In what follows in this section we will assume ħ=1, $\omega_k^0 = \omega_0 = 1, M_e = 1$. It follows from (22) that the difference between the bipolaron Hamiltonian and the polaron one is that in the latter $V_k$ is replaced by $\bar{V}_k$ and $\bar{T}, \bar{U}, \bar{\Pi}$ are added to the polaron Hamiltonian.

With the use of the Lee-Low-Pines canonical transformation:

$$S_2 = \exp\left\{\sum_k f_k(a_k^+ - a_k)\right\}, \tag{24}$$

where $f_k$are variational parameters which stand for the value of the displacement of the field oscillators from their equilibrium positions:

$$S_2^{-1}a_k S_2 = a_k + f_k, \qquad S_2^{-1}a_k^+ S_2 = a_k^+ + f_k, \tag{25}$$

Hamiltonian $\tilde{\tilde{H}}$:

$$\tilde{\tilde{H}} = S_2^{-1}\tilde{H}S_2, \tag{26}$$

$$\tilde{\tilde{H}} = H_0 + H_1,$$

will be written as:

$$H_0 = 2\sum_k \bar{V}_k f_k + \sum_k f_k^2 \widetilde{\omega}_k + \frac{1}{2}\left(\sum_k \boldsymbol{k} f_k^2\right)^2 + \mathcal{H}_0 + \bar{T} + \bar{U} + \bar{\Pi},$$

$$\mathcal{H}_0 = \sum_k \omega_k\, a_k^+ a_k + \frac{1}{2}\sum_{k,k'} \boldsymbol{k}\boldsymbol{k}'\, f_k f_{k'}\left(a_k a_{k'} + a_k^+ a_{k'}^+ + a_k^+ a_{k'} + a_{k'}^+ a_k\right), \tag{27}$$

where:

$$\omega_k = \widetilde{\omega}_k + \frac{k^2}{2} + \boldsymbol{k}\sum_{k'} \boldsymbol{k}'\, f_{k'}^2. \tag{28}$$

Hamiltonian $H_1$ contains the terms which are linear, three-fold and four-fold in the creation and annihilation operators. Its explicit form is given in [44], [75] (supplement).

Then, according to [44], [75], the Bogolyubov-Tyablikov canonical transformation [112] is used to pass on from the operators $a_k^+, a_k$ to $\alpha_k^+, \alpha_k$:

$$a_k = \sum_{k'} M_{1kk'}\, \alpha_{k'} + \sum_{k'} M_{2kk'}^* \alpha_{k'}^+$$

$$a_k^+ = \sum_{k'} M_{1kk'}^* \alpha_{k'}^+ + \sum_{k'} M_{2kk'} \alpha_{k'}, \tag{29}$$

where $\mathcal{H}_0$ is a diagonal operator which makes vanish expectation $H_1$ in the absence of an external magnetic field (see Supplement). The contribution of $H_1$ into the spectrum of the transformed Hamiltonian when the magnetic field is non-zero is discussed in Section 5.

In the new operators $\alpha_k^+, \alpha_k$ Hamiltonian (27) takes on the form:

$$\tilde{\tilde{\tilde{H}}} = E_{bp} + \sum_k v_k\, \alpha_k^+ \alpha_k,$$

$$E_{bp} = \triangle E_r + 2\sum_k \bar{V}_k f_k + \sum_k \widetilde{\omega}_k f_k^2 + \bar{T} + \bar{U} + \bar{\Pi}, \tag{30}$$

where $\triangle E_r$ is the so called “recoil energy”. A general expression for $\triangle E_r = \triangle E_r\{f_k\}$ was obtained in [75]. The ground state energy $E_{bp}$was calculated in [69]-[71] by minimization of (30) with respect to $f_k$ and $\psi$ in the absence of a magnetic field.

It should be noted that in the polaron theory with a broken symmetry a diagonal electron-phonon Hamiltonian takes the form (30) [113]. This Hamiltonian can be interpreted as a Hamiltonian of a polaron and a system of its associated renormalized actual phonons or a Hamiltonian which possesses a spectrum of quasiparticle excitations determined by (30) [114]. In the latter case the polaron excited states are Fermi quasiparticles.

In the case of a bipolaron the situation is qualitatively different because a bipolaron is a Bose particle whose spectrum is determined by (30). Obviously, a gas of such particles can experience Bose-Einstein condensation. Treatment of (30) as a bipolaron and its associated renormalized phonons does not prevent their BEC, since maintenance of the particles required for BEC is fulfilled automatically since the total number of the renormalized phonons commutate with Hamiltonian (30).

Renormalized frequencies $\nu_k$ involved in (30), according to [44], [75], [115], are determined by a secular equation for $s$:

$$1 = \frac{2}{3}\sum_k \frac{k^2 f_k^2 \omega_k}{s-\omega_k^2}, \tag{31}$$

whose solutions yield a spectrum of the values of $s = \{\nu_k^2\}$.

**5. Energy spectrum of a TI bipolaron.**

Hamiltonian (30) can be conveniently presented as:

$$\tilde{\tilde{\tilde{H}}} = \sum_{n=0,1,2\dots} E_n \alpha_n^+ \alpha_n\,, \tag{32}$$

$$E_n = \begin{Bmatrix} E_{bp\,,} \; n=0; \\ \nu_n = E_{bp} + \omega_{k_n}\,, \;\; n \neq 0; \end{Bmatrix}, \tag{33}$$

where, in the case of a three-dimensional ionic crystal, $\vec{k}_n$ is a vector with the components:

$$k_{n_i} = \pm\frac{2\pi(n_i-1)}{N_{a_i}}, \qquad n_i = 1,2,\dots,\frac{N_{a_i}}{2}+1, \qquad i = x,y,z, \tag{34}$$

$N_{a_i}$ is the number of atoms along the $i$ -th crystallographic axis. Let us prove the validity of the expression for the spectrum (32), (33). Since the operators $\alpha_n^+$, $\alpha_n$ obey Bose commutation relations:

$$[\alpha_n, \alpha_{n'}^+] = \alpha_n\alpha_{n'}^+ - \alpha_{n'}^+\alpha_n = \delta_{n,n'}, \tag{35}$$

they can be considered to be operators of the birth and annihilation of TI bipolarons. The energy spectrum of TI bipolarons, according to (31), is given by the equation:

$$F(s) = 1, \tag{36}$$

where:

$$F(s) = \frac{2}{3}\sum_n \frac{k_n^2 f_{k_n}^2 \omega_{k_n}^2}{s-\omega_{k_n}^2}.$$

It is convenient to solve equation (36) graphically (Fig.1):

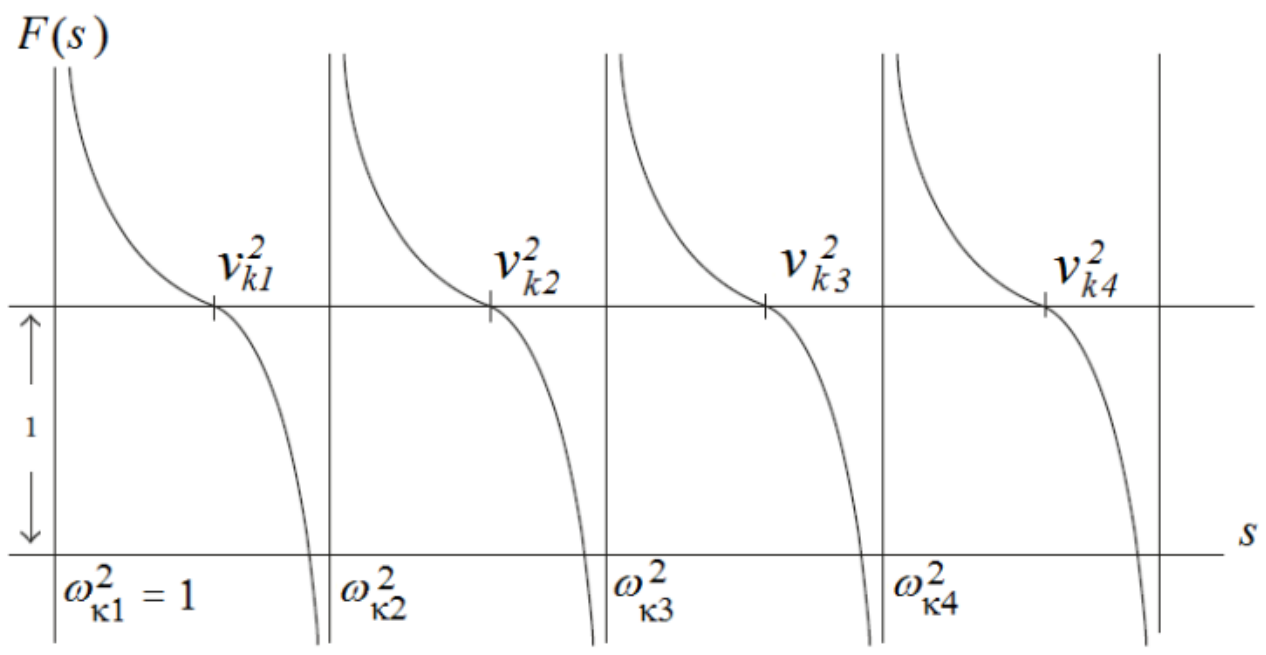

Fig.1. Graphical solution of equation (36).

Fig.1. suggests that the frequencies $v_{k_n}$ lie between the frequencies $\omega_{k_n}$ and $\omega_{k_{n+1}}$. Hence, the spectrum $v_{k_n}$ as well as the spectrum $\omega_{k_n}$ is quasicontinuous: $v_{k_n} - \omega_{k_n} = 0(N^{-1})$, which just proves the validity of (32), (33).

It follows that the spectrum of a TI bipolaron has a gap between the ground state $E_{bp}$ and a quasicontinuous spectrum equal to $\omega_0$.

In the absence of a magnetic field functions $f_k$ involved in the expression for $\omega_k$ (28) are independent of the direction of the wave vector $\vec{k}$. If a magnetic field is present, $f_k$ cannot be considered to be an isotropic value, accordingly, the last term in expression (28) for $\omega_k$ cannot be assumed to be zero. Besides, an angular dependence involved the spectrum $\omega_k$ in a magnetic field is also contained in the term $\widetilde{\omega}_k$ involved in $\omega_k$. Since in the isotropic system under consideration there is only one preferred direction determined by vector $\vec{B}$, for $\omega_k$ from (28), we get:

$$\omega_{k_n} = \omega_0 + \frac{\hbar^2 k_n^2}{2M_e} + \frac{\eta}{M_e}(\boldsymbol{B}\boldsymbol{k}_n), \tag{37}$$

where η is a certain scalar value. It should be noted that the contribution $H_1$ into spectrum (37) leads to a dependence of η on $(\vec{k}\vec{B})$ too. For a weak magnetic field in a long-wave limit (when Fröhlich Hamiltonian is applicable) we can neglect this dependence and believe that η is constant.

For a magnetic field directed along the z-axis, expression (37) can be presented in the form:

$$\omega_{k_n} = \omega_0 + \frac{\hbar^2}{2M_e}(k_{zn} + k_z^0)^2 + \frac{\hbar^2}{2M_e}\left(k_{xn}^2 + k_{yn}^2\right) - \frac{\eta^2 B^2}{2\hbar^2 M_e}. \tag{38}$$

It should be noted that formula (38) can be generalized to the anisotropic case (which will be actual in what follows) when in the directions $k_x$ and $k_y$: $M_{ex} = M_{ey} = M_{||}$, and in the direction $k_z$: $M_{ez} = M_{\perp}$ (Section 8).

In this case formula (38) takes the form:

$$\omega_{k_n} = \omega_0 + \frac{\hbar^2}{2M_{\perp}}(k_{zn} + k_z^0)^2 + \frac{\hbar^2}{2M_{||}}\left(k_{xn}^2 + k_{yn}^2\right) - \frac{\eta^2 B^2}{2\hbar^2 M_{\perp}}, \tag{38'}$$

if a magnetic field is directed along the z-axis and:

$$\omega_{k_n} = \omega_0 + \frac{\hbar^2}{2M_\perp} k_{zn}^2 + \frac{\hbar^2}{2M_{||}} (k_{xn} + k_x^0)^2 - \frac{\hbar^2}{2M_{||}} k_{yn}^2 - \frac{\eta^2 B^2}{2\hbar^2 M_{||}}, \qquad (38'')$$

if a magnetic field is directed along the x-axis.

Below we consider the case of a low concentration of TI bipolarons in a crystal. In this case, as we will show in the next section, they can be considered as an ideal Bose gas whose properties are determined by Hamiltonian (32).

## 6. Nonideal gas of TI bipolarons

Being charged, a gas of TI bipolarons cannot be ideal since there should be a Coulomb interaction between the polarons. The theory of a nonideal gas implies that consideration of an interaction between the particles leads to qualitative changes in its spectral properties. According to [101], consideration of even a short-range interaction leads to a gap in the spectrum which is lacking in an ideal gas. Even more considerable changes can be expected in the case of a long-range Coulomb interaction. In this section we restrict ourselves by a lack of a magnetic field.

The logical scheme of the approach is as follows:

a) first we consider a particular case when there are only two electrons interacting with a phonon field. This is a classical bipolaron problem [44];

b) then we deal with a multiparticle problem which leads to the Fermi liquid concept. For this multielectron problem, we consider the case of two additional electrons occurring above the Fermi surface (in its vicinity) bound by EPI (Cooper problem) [116];

c) then we believe that nearly all the electrons lying in the energy level within the layer $[E_F + E_{pol}, E_F]$, where $E_F$ – is the Fermi energy, $E_{pol}$ – is the polaron energy, occur in the TI polaron state; accordingly all the electrons in the narrow layer $[E_F + E_{bp}/2 - \delta E, E_F + E_{bp}/2 + \delta E], \delta E \to 0$, occur in the TI bipolaron state, where $E_{bp}$ is the energy of a TI bipolaron. A condensed bipolaron gas leads to an infinite density of electron states in this level;

d) bipolarons are considered as charged bosons placed in an electron Fermi liquid (polaron gas) which screens an interaction between the bipolarons and the problem is reduced to that of a non-ideal charged Bose gas;

e) the spectrum obtained is used to calculate the statistical properties of a TI bipolaron gas.

To develop a theory of a non-ideal TI bipolaron gas we should know the spectrum of the states of an individual TI bipolaron in a polar medium. This problem was considered in detail in [117]-[118] (Section 5). As it is shown in [116], this spectrum will be the same as that of TI bipolarons which emerge near the Fermi surface. Hence, TI bipolarons in the layer $[E_F + E_{bp}/2, E_F]$ can be considered as a TI bipolaron Bose gas occurring in a polaron gas [119]. If we believe that TI bipolarons do not interact with one another, such a gas can be considered to be ideal. Its properties will be fully determined if we know the spectrum of an individual TI bipolaron.

In considering the theory of an ideal gas and superconductivity on the basis of Bose particles of TI bipolarons, the Coulomb interaction between the electrons is taken into account only for electron pairs, i.e. when we deal with the problem of an individual bipolarons. Hamiltonian of such a system, according to [44], [118] has the form:

$$H_0 = \sum_k \varepsilon_k \alpha_k^+ \alpha_k, \qquad (39)$$

$$\varepsilon_k = E_{bp}\Delta_{k,0} + \left(\omega_0 + E_{bp} + k^2/2M_e\right)\left(1 - \Delta_{k,0}\right), \qquad (40)$$

where $\alpha_k^+, \alpha_k$ are operators of the birth and annihilation of TI bipolarons: $\varepsilon_k$ is the spectrum of TI bipolarons obtained in Section 5;
$\omega_0(\vec{k}) = \omega_0$ is the energy of an optical phonon; $\Delta_{k,0}= 1$ for $k = 0$ and $\Delta_{k,0}= 0$ for $k \neq 0$. Expression (39), (40) can be rewritten as:

$$H_0 = E_{bp}\alpha_0^+\alpha_0 + \sum_k{}' (\omega_0 + E_{bp} + k^2/2M_e)\alpha_k^+ \alpha_k, \tag{41}$$

where the prime in the sum in the right-hand side of (41) means that the term with $k = 0$ is lacking in the sum. Extraction out of a term with $k = 0$ in (41) corresponds to formation of a Bose condensate where:

$$\alpha_0 = \sqrt{N_0}\ , \tag{42}$$

$N_0$ is the number of TI bipolarons in a condensed state. Hence in the theory of an ideal TI bipolaron gas the first term is merely $E_{bp}N_0$. In constructing a theory of a non-ideal TI bipolaron Bose gas we will proceed from the Hamiltonian:

$$H = E_{bp}N_0. + \sum_k{}' (\omega_0 + E_{bp})\alpha_k^+ \alpha_k + \sum_k{}' t_k\alpha_k^+ \alpha_k + 1/2V \sum_k{}' V_k\alpha_{k''-k}^+\alpha_{k'+k}^+\alpha_{k''}\alpha_{k'},$$

$$t_k = k^2/2M_e\ , \tag{43}$$

where the last term responsible for bipolaron interaction is added to Hamiltonian $H_0$ (41),$V_k$is a matrix element of the bipolaron interaction. The last two terms in (43) exactly correspond to Hamiltonian of a charged Bose gas [120]. Following a standard procedure of resolving a Bose condensate we rewrite (43) into the Hamiltonian:

$$H = E_{bp}N_0 + \sum_k{}' (\omega_0 + E_{bp})\,\alpha_k^+\alpha_k + \tag{44}$$

$$\sum_k{}' [(t_k + n_0V_k)\alpha_k^+\alpha_k + 1/2n_0V_k(\alpha_k\alpha_{-k} + \alpha_k^+\alpha_{-k}^+)],$$

where $n_0 = N_0/V$ is the density of the particles in the Bose condensate.
Then with the use of Bogolyubov transformation:

$$\alpha_k = u_kb_k - v_kb_{-k}^+, \tag{45}$$

$$u_k = [(t_k + n_0V_k + \epsilon_k)/2\epsilon_k]^{1/2},$$

$$v_k = [(t_k + n_0V_k - \epsilon_k)/2\epsilon_k]^{1/2}, \qquad \epsilon_k = [2n_0V_kt_k + t_k^2]^{1/2}$$

in new operators we get the Hamiltonian

$$H = E_{bp}N_0 + U_0 + \sum_k{}' (\omega_0 + E_{bp} + \epsilon_k)b_k^+b_k, \tag{46}$$

$$U_0 = \sum_k{}' (\epsilon_k - t_k - n_0V_k),$$

where $U_0$is the ground state energy of a charged Bose gas without regard to its interaction with the crystal polarization. Hence, the excitation spectrum of a non-ideal TI bipolaron gas has the form:

$$E_k = E_{bp} + u_0 + \left(\omega_0(\boldsymbol{k}) + \sqrt{k^4/4M_e^2 + k^2 V_k n_0/M_e}\right)\left(1 - \Delta_{k,0}\right), \qquad (47)$$

where $u_0 = U_n/N$, $N$ is the total number of particles. If we reckon the excitation energy from the bipolaron ground state energy in a non-ideal gas, assuming that $\Delta_k = E_k - \left(E_{bp} + u_0\right)$, then for $\Delta_k$ , (when $k \neq 0$ ) we get:

$$\Delta_k = \omega_0(\boldsymbol{k}) + \sqrt{k^4/4M_e^2 + k^2 V_k n_0/M_e}. \qquad (48)$$

This spectrum suggests that a TI bipolaron gas has a gap $\Delta_k$ in the spectrum between the ground and excited states, i.e. is superfluid. Being charged, this gas is automatically superconducting. To determine a particular form of the spectrum we should know the value of $V_k$. If we considered only a charged Bose gas with a positive homogeneous background, produced by a rigid ion lattice, then the quantity $V_k$ involved in (47) in the absence of screening would be equal to $V_k = 4\pi e_B^2/k^2$. Accordingly the second term in the radical expression in (47) would be equal to $\omega_p^2 = 4\pi e_B^2/M_e$, where $\omega_p$ is the plasma frequency of a Boson gas, $e_B$is the Boson charge ($2e$ for a TI bipolaron). Actually, if screening is taken into account, $V_k$ takes the form $V_k = 4\pi e^2/k^2\epsilon_B(k)$, where $\epsilon_B(k) -$ is the dielectric permittivity of a charged Bose gas which was calculated in [121],[122]. The expression for $\epsilon_B(k)$ obtained in [121],[122] is too cumbersome and is not given here. However, in the case of a TI bipolaron Bose gas this modification of $V_k$, is insufficient. As it was shown in [116] (Section 7), bipolarons constitute just a small portion of charged particles in the system. Most of them occur in the electron gas into which the bipolarons are immersed. It is just the electron gas that makes the main contribution into the screening of the interaction between the polarons. To take account of this screening, $V_k$ should be expressed as $V_k = 4\pi e^2/k^2\epsilon_B(k)\epsilon_e(k)$, where $\epsilon_e(k)$,is the dielectric permittivity of an electron gas. Finally, if we take account of the mobility of the ion lattice, $V_k$ takes the form: $V_k = 4\pi e^2/k^2\epsilon_B(k)\epsilon_e(k)\epsilon_\infty\epsilon_0$ , where $\epsilon_\infty, \epsilon_0$ are the high-frequency and static dielectric constants.

As a result we get for $\Delta_k$:

$$\Delta_k = \omega_0(\boldsymbol{k}) + k^2/2M_e\sqrt{1 + \chi(k)}, \qquad (49)$$

$$\chi(k) = \left(2M_e\omega_p\right)^2/k^4\epsilon_B(k)\epsilon_e(k)\epsilon_\infty\epsilon_0. \qquad (50)$$

To estimate $\chi(k)$ in (49) let us consider the long-wave limit. In this limit $\epsilon_e(k)$ has the Thomas-Fermi form: $\epsilon_e(k) = 1 + \varkappa^2/k^2$, where $\varkappa = 0{,}815 k_F (r_s/a_B)^{1/2}$, $a_B = \hbar/M_e e_B^2$, $r_s = (3/4\pi n_0)^{1/3}$, therefore, according to [121], [122], the quantity $\epsilon_B(k)$ is equal to: $\epsilon_B(k) = 1 + q_s^4/k^4, q_s = \sqrt{2M_e\omega_p}$.

Bearing in mind that in calculations of the thermodynamic functions the main contribution is made by the values of $k$: $k^2/2M_e \approx T$, where $T$ is the temperature for $\chi(k)$, we will get an estimate $\chi \sim T/E_F\epsilon_\infty\epsilon_0$ where $E_F$ –is the Fermi energy. Hence the spectrum of a screened TI bipolaron gas differs from the spectrum of an ideal bipolaron gas (40) only slightly. It should be noted that in view of screening the value of the correlation energy $u_0$ in (48) turns out to be much less than that calculated in [120] without screening and for actual parameter values – much less than the bipolaron energy $|E_{bp}|$. It should also be noted that in view of screening a TI bipolaron gas does not form Wigner crystal even in the case of an arbitrarily small bipolaron density.

**7. Statistical thermodynamics of a low-density TI bipolaron gas.**

In accordance with the result of the previous section let us consider an ideal Bose gas of TI bipolarons which is a system of N particles occurring in a volume V [117]-[118]. Let us write $N_0$ for the number of particles in the lower one-particle state, and $N'$– for the number of particles in higher states. Then:

$$N=\sum_{n=0,1,2,\ldots}\bar{m}_n=\sum_n\frac{1}{e^{(E_n-\mu)/T}-1}, \tag{51}$$

or:

$$N=N_0+N',\ \ N_0=\frac{1}{e^{\frac{(E_{bp}-\mu)}{T}}-1},$$

$$N'=\sum_{n\neq 0}\frac{1}{e^{(E_n-\mu)/T}-1}\ . \tag{52}$$

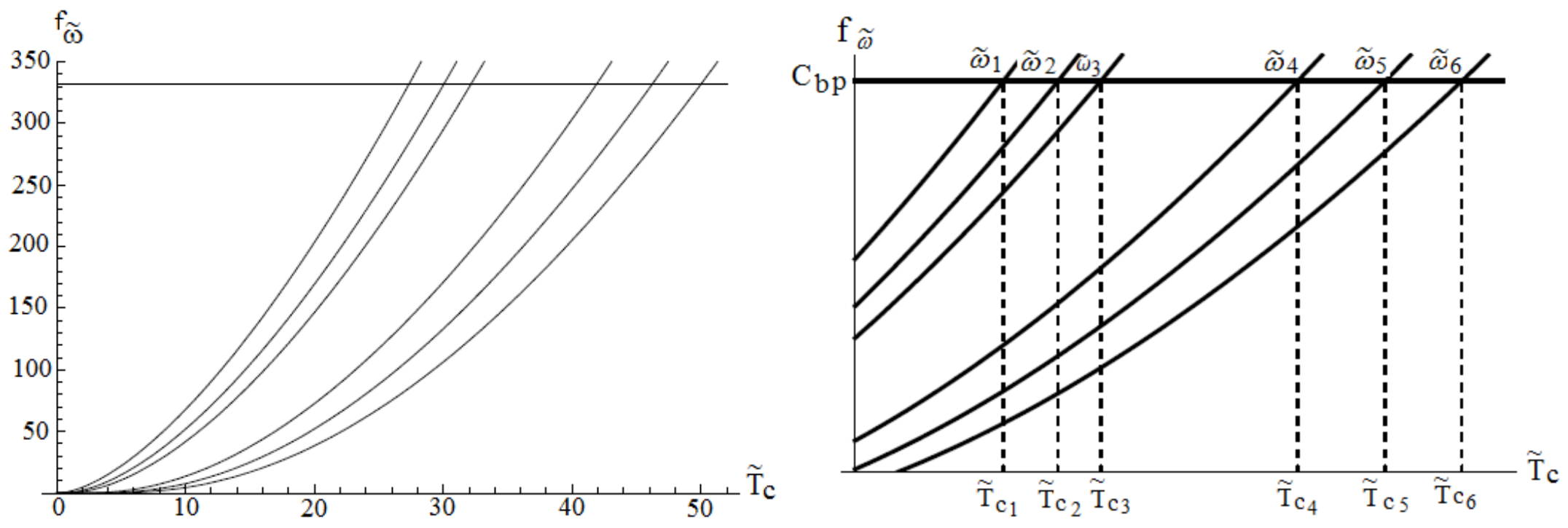

Fig.2 Solutions of equation (53) with $C_{bp}=331.3$ and $\tilde{\omega}_i=\{0.2;1;2;10;15;20\}$, which correspond to $\tilde{T}_{c_i}$: $\tilde{T}_{c_1}=27.3$; $\tilde{T}_{c_2}=30$; $\tilde{T}_{c_3}=32$; $\tilde{T}_{c_4}=42$; $\tilde{T}_{c_5}=46.2$; $\tilde{T}_{c_6}=50$.

If in the expression for $N'$ (52) we replace summation by integration over a continuous spectrum (32), (33),(38) and assume in (52) that $\mu=E_{bp}$, we will get from (51), (52) an equation for the critical temperature of Bose condensation $T_c$:

$$C_{bp}=f_{\tilde{\omega}_H}(\tilde{T}_c), \tag{53}$$

$$f_{\tilde{\omega}_H}(\tilde{T}_c)=\tilde{T}_c^{3/2}F_{3/2}(\tilde{\omega}_H/\tilde{T}_c),\qquad F_{3/2}(\alpha)=\frac{2}{\sqrt{\pi}}\int_0^\infty\frac{x^{1/2}dx}{e^{x+\alpha}-1},$$

$$C_{bp}=\left(\frac{n^{2/3}2\pi\hbar^2}{M_e\omega^*}\right)^{3/2},\tilde{\omega}_H=\frac{\omega_0-\eta^2H^2/2M_e}{\omega^*},\qquad \tilde{T}_c=\frac{T_c}{\omega^*},$$

where n=$N/V$. In this section we deal with the case when a magnetic field is absent: $H$=0. Fig.2 shows a graphical solution of equation (53) for the parameter values $M_e = 2m^* = 2m_0$, where $m_0$ is a mass of a free electron in vacuum,$\omega^* = 5\text{meV}(\approx 58\text{K}), n = 10^{21}\text{cm}^{-3}$ and the values: $\tilde{\omega}_1 = 0.2$; $\tilde{\omega}_2 = 1$; $\tilde{\omega}_3 = 2$; $\tilde{\omega}_4 = 10$; $\tilde{\omega}_5 = 15$; $\tilde{\omega}_6 = 20$; $\tilde{\omega}_H = \tilde{\omega} = \frac{\tilde{\omega}_0}{\omega^*}, \omega_{H_{cr}} = \omega_H$ for $\tilde{T} = \tilde{T}_C$.

Fig.2 suggests that the critical temperature grows as the phonon frequency $\tilde{\omega}_0$ increases. The ratios of the critical temperatures $T_{ci}/\omega_{0i}$ corresponding to these parameter values are given in Table 1. It is evident from Table 1 that the critical temperature of a TI bipolaron gas is always higher than that in the case of an ideal Bose gas (IBG). Fig.2 also suggests that an increase in the concentration of TI bipolarons $n$ will lead to an increase in the critical temperature, while an increase in the electron mass $m^*$ – to its decrease.

For $\tilde{\omega} = 0$ the results go over in to the limit of IBG. In particular, from (53) for $\tilde{\omega} = 0$ it follows that the critical temperature of IBG is:

$$T_c = 3.31\hbar^2 n^{2/3}/M_e. \tag{54}$$

Table 1. Calculated values of characteristics of a TI bipolaron Bose gas with concentration $n = 10^{21}\text{cm}^{-3}$. $\tilde{\omega}_i = \frac{\omega_i}{\omega^*}, \omega^* = 5\text{meV}, \omega_i$ – is the energy of an optical phonon; $T_{c_i}$ –is a critical temperature of the transition, $q_i$ is a latent heat of the transition from a condensate to a supracondensate state; $-\Delta(\partial C_{vi}/\partial\tilde{T}) = \partial C_{vi}/\partial\tilde{T}|_{\tilde{T}=T_{c_i}+0} - \partial C_{vi}/\partial\tilde{T}|_{\tilde{T}=T_{c_i}-0}$ is a jump of the heat capacity during a SC transition,$\tilde{T} = T/\omega^*$; $C_{v,i}(T_c - 0)$is the heat capacity in the SC phase at the critical point; $C_s = C_v(T_c - 0)$, $C_n = C_v(T_c + 0)$. Calculations are performed for the concentration of TI bipolarons $n = 10^{21}\text{cm}^{-3}$ and the effective mass of a band electron $m^* = m_0$. The table also lists the values of the concentrations of TI bipolarons $n_{bpi}$ for HTSC $Y\, B_{a_2} C_{u_3} O_7$, proceeding from the experimental value of the transition temperature $T_c$= 93$K$.

| $i$ | 0 | 1 | 2 | 3 | 4 | 5 | 6 |
|---|---|---|---|---|---|---|---|
| $\tilde{\omega}_i$ | 0 | 0.2 | 1 | 2 | 10 | 15 | 20 |
| $T_{ci}/\omega_{0i}$ | ∞ | 136.6 | 30 | 16 | 4.2 | 3 | 2.5 |
| $q_i/T_{ci}$ | 1.3 | 1.44 | 1.64 | 1.8 | 2.5 | 2.8 | 3 |
| $-\Delta(\partial C_{vi}/\partial\tilde{T})$ | 0.11 | 0.12 | 0.12 | 0.13 | 0.14 | 0.15 | 0.15 |
| $C_{vi}(T_c - 0)$ | 1.9 | 2.16 | 2.46 | 2.7 | 3.74 | 4.2 | 1,6 |
| $(C_s - C_n)/C_n$ | 0 | 0.16 | 0.36 | 0.52 | 1.23 | 1.53 | 1.8 |
| $n_{bp_i} \cdot cm^3$ | $16 \cdot 10^{19}$ | $9.4 \cdot 10^{18}$ | $4.2 \cdot 10^{18}$ | $2.0 \cdot 10^{18}$ | $1.2 \cdot 10^{17}$ | $5.2 \cdot 10^{14}$ | $2.3 \cdot 10^{13}$ |

It should be stressed that (54) involves $M_e = 2m$, rather than the bipolaron mass. This eliminates the problem of low condensation temperature which arises both in the SRP and LRP theories where expression (54) involves the bipolaron mass [37], [58], [123]-[128]. Another important result is that the critical temperature $T_c$ for the parameter values chosen is considerably superior to the gap energy $\omega_0$.

From (51), (52) it follows that:

$$\frac{N'(\widetilde{\omega})}{N}=\frac{\tilde{T}^{\frac{3}{2}}}{C_{bp}}F_{\frac{3}{2}}\left(\frac{\widetilde{\omega}}{\tilde{T}}\right), \tag{55}$$

$$\frac{N_0(\widetilde{\omega})}{N}=1-\frac{N'(\widetilde{\omega})}{N}. \tag{56}$$

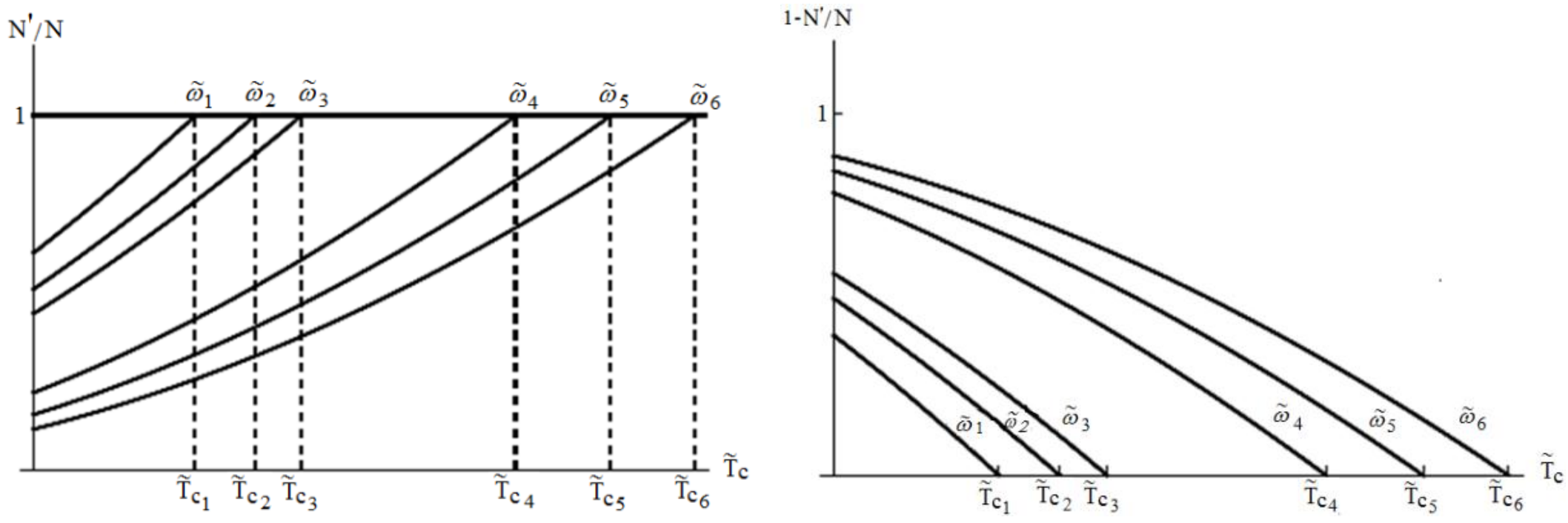

Fig. 3. Temperature dependencies of the relative number of supracondensate $N'/N$ and condensate $N_0/N=1-N'/N$, particles for the values of the parameter $\widetilde{\omega}_i$, presented in Fig.2.

Fig.3 shows temperature dependencies of the number of supracondensate $N'$ and condensate $N_0$ particles for the above indicated values of $\widetilde{\omega}_i$.

Fig.3 suggests that, as one would expect, the number of condensate particles grows as the gap $\widetilde{\omega}_i$increases.

The energy $E$ of a TI bipolaron gas is determined by the expression:

$$E=\sum_{n=0,1,2\dots}\bar{m}_nE_n=E_{bp}N_0+\sum_{n\neq 0}\bar{m}_nE_{n.} \tag{57}$$

With the use of (32), (33) and (57) we express the specific energy (i.e. energy per one TI bipolaron) $\tilde{E}(\tilde{T})=E/N\omega^*$, $\tilde{E}_{bp}=E_{bp}/\omega^*$ as:

$$\tilde{E}(\tilde{T})=\tilde{E}_{bp}+\frac{\tilde{T}^{5/2}}{C_{bp}}F_{3/2}\left(\frac{\widetilde{\omega}-\tilde{\mu}}{\tilde{T}}\right)\left[\frac{\widetilde{\omega}}{\tilde{T}}+\frac{F_{5/2}\left(\frac{\widetilde{\omega}-\tilde{\mu}}{\tilde{T}}\right)}{F_{3/2}\left(\frac{\widetilde{\omega}-\tilde{\mu}}{\tilde{T}}\right)}\right], \tag{58}$$

$$F_{5/2}(\alpha)=\frac{2}{\sqrt{\pi}}\int_0^\infty\frac{x^{3/2}dx}{e^{x+\alpha}-1},$$

where $\tilde{\mu}$ is determined by the equation:

$$\tilde{T}^{3/2}F_{3/2}\left(\frac{\tilde{\omega}-\tilde{\mu}}{\tilde{T}}\right)=C_{bp}\ , \qquad (59)$$

$$\tilde{\mu}=\begin{cases}0, & \tilde{T}\le\tilde{T}_c;\\ \tilde{\mu}(\tilde{T}), & \tilde{T}\ge\tilde{T}_c\end{cases}.$$

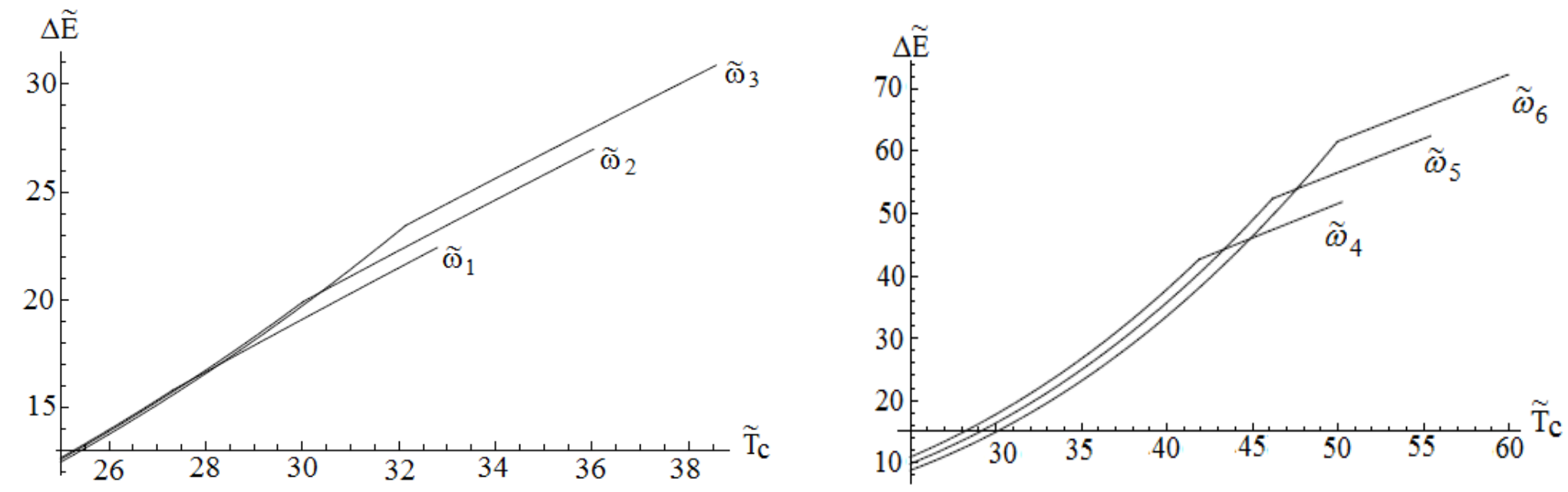

Fig. 4 Temperature dependencies of $\Delta E(\tilde{T})=\tilde{E}(\tilde{T})-\tilde{E}_{bp}$ for $\tilde{\omega}_i$ values presented in Fig.2,3

A relation of $\tilde{\mu}$ to the chemical potential of the system μ is determined by the expression $\tilde{\mu}=(\mu-E_{bp})/\omega^*$. Formulae (58)-(59) also yield the expressions for the free energy *F*=-2*E/3* and entropy *S*=-∂*F/*∂*T.*

Fig.4 shows temperature dependencies of $\Delta E=\tilde{E}-\tilde{E}_{bp}$ for the above-given values of $\tilde{\omega}_i$. Breakpoints of the $\Delta E_i(\tilde{T})$ curves correspond to the critical temperature values $T_{ci}$.

The dependencies obtained enable us to find the heat capacity of a TI bipolaron gas: $C_v(\tilde{T})=d\tilde{E}/d\tilde{T}$. With the use of (58) for $\tilde{T}\le\tilde{T}_c$ we express $C_v(\tilde{T})$ as:

$$C_v(\tilde{T})=\frac{\tilde{T}^{3/2}}{2C_{bp}}\left[\frac{\tilde{\omega}^2}{\tilde{T}^2}F_{1/2}\left(\frac{\tilde{\omega}}{\tilde{T}}\right)+6\left(\frac{\tilde{\omega}}{\tilde{T}}\right)F_{3/2}\left(\frac{\tilde{\omega}}{\tilde{T}}\right)+5F_{5/2}\left(\frac{\tilde{\omega}}{\tilde{T}}\right)\right], \qquad (60)$$

$$F_{\frac{1}{2}}(\alpha)=\frac{2}{\sqrt{\pi}}\int_0^\infty\frac{1}{\sqrt{x}}\frac{dx}{e^{x+\alpha}-1}.$$

Expression (60) yields a well-known exponential dependence of the heat capacity at low temperatures $C_v\sim\exp(-\omega_0/T)$, caused by the availability of the energy gap $\omega_0$.

Fig.5 illustrates the temperature dependencies of the heat capacity $C_v(\tilde{T})$ for the above-mentioned values of $\tilde{\omega}_i$. Table 1 lists the values of jumps in the heat capacity for different $\tilde{\omega}_i$:

$$\Delta\frac{\partial C_v(\tilde{T})}{\partial\tilde{T}}=\frac{\partial C_v(\tilde{T})}{\partial\tilde{T}}\Big|_{\tilde{T}=\tilde{T}_c+0}-\frac{\partial C_v(\tilde{T})}{\partial\tilde{T}}\Big|_{\tilde{T}=\tilde{T}_c-0} \qquad (61)$$

at the transition points.

The dependencies obtained enable one to find a latent transition heat *q*=*TS*, *S* is the entropy of supracondensate particles. At the transition point this value is equal to:

$q=2T_cC_v(T_C-0)/3$,, where $C_v(T)$ is determined by formula (60), and for the above-mentioned values of $\tilde{\omega}_i$, is given in Table1

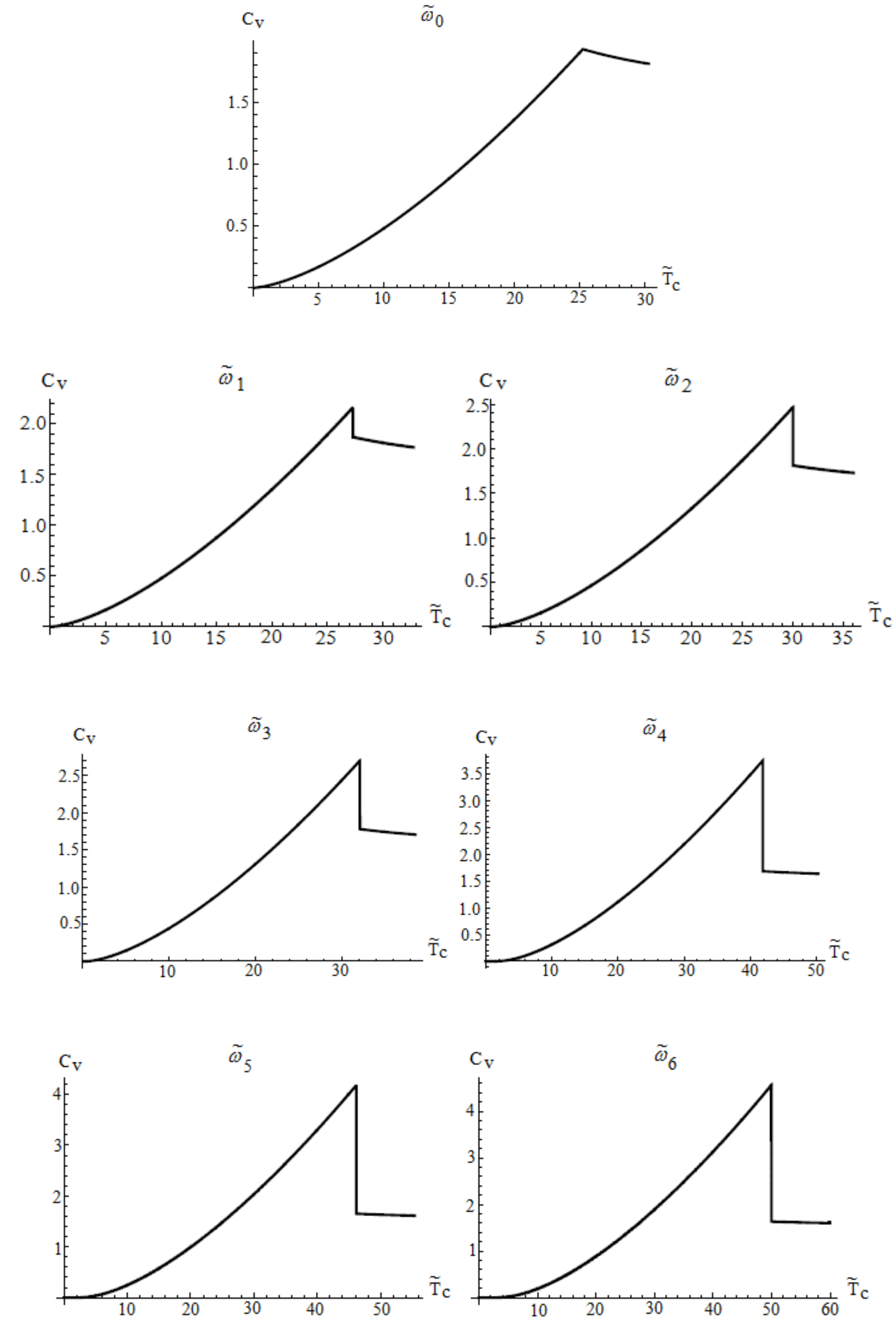

Fig. 5. Temperature dependencies of the heat capacity for different values of the parameters $\widetilde{\omega}_i$: $\omega_0$=0; $\tilde{T}_C = 25.2$; $C_v(\tilde{T}_{c0}) = 1.9$;

$$\omega_1 = 0.2;\ \tilde{T}_{c_1} = 27.3;\ C_v(\tilde{T}_{c1} - 0) = 2.16;\ C_v(\tilde{T}_{c1} + 0) = 1.9;$$
$$\omega_2 = 1;\ \tilde{T}_{c_2} = 30;\ C_v(\tilde{T}_{c2} - 0) = 2.46;\ C_v(\tilde{T}_{c2} + 0) = 1.8;$$
$$\omega_3 = 2;\ \tilde{T}_{c_3} = 32.1;\ C_v(\tilde{T}_{c3} - 0) = 2.7;\ C_v(\tilde{T}_{c3} + 0) = 1.78;$$
$$\omega_4 = 10;\ \tilde{T}_{c_4} = 41.9;\ C_v(\tilde{T}_{c4} - 0) = 3.7;\ C_v(\tilde{T}_{c4} + 0) = 1.7;$$
$$\omega_5 = 15;\ \tilde{T}_{c_5} = 46.2;\ C_v(\tilde{T}_{c5} - 0) = 4.2;\ C_v(\tilde{T}_{c5} + 0) = 1.65;$$
$$\omega_6 = 20;\ \tilde{T}_{c_6} = 50;\ C_v(\tilde{T}_{c6} - 0) = 4.6;\ C_v(\tilde{T}_{c6} + 0) = 1.6.$$

**8. Current states of a TI bipolaron gas.**

As is well known, an absence of a magnetic field in a superconductor is caused by a presence of surface currents which compensate this field. Thus, it follows from condition (20) that:

$$\boldsymbol{P}_R = -\frac{2e}{c}\boldsymbol{A}_R, \tag{62}$$

i.e. in the superconductor there is a persistent current $\vec{\boldsymbol{j}}$**:**

$$\boldsymbol{j} = 2en_0\boldsymbol{P}_R/M_e^* = -\frac{4e^2n_0}{M_e^*c}\boldsymbol{A}_R \tag{63}$$

(where $M_e^*$ is the bipolaron effective mass) which leads to Meissner effect where $n_0$ is the concentration of superconducting current carriers: $n_0 = N_0/V$. Comparing (63) with the well-known phenomenological expression for a surface current $\vec{j}_S$ [129]:

$$\boldsymbol{j}_S = -\frac{c}{4\pi\lambda^2}\boldsymbol{A}, \tag{64}$$

and assuming that $\vec{A} = \vec{A}_R$, from equality $\vec{j} = \vec{j}_S$ and (63), (64) we derive a well-known expression for the London penetration depth λ:

$$\lambda = \left(\frac{M_e^*c^2}{16\pi e^2n_0}\right)^{1/2}. \tag{65}$$

The equality of “microscopic” current expression (63) to its “macroscopic” value cannot be exact. Accordingly, the equality $\vec{A} = \vec{A}_R$ is also approximate since $\vec{A}_R$ is a vector-potential at the point where the mass center of two electrons occurs, while in the London theory $\vec{A}$ is a vector-potential at the point where the particle occurs. Therefore, it would be more realistic to believe that these quantities are proportional. In this case the expression for the penetration depth will be:

$$\lambda = \mathrm{const}\left(\frac{M_e^*c^2}{16\pi e^2n_0}\right)^{1/2}, \tag{65$'$}$$

where the constant multiplier (of the order of 1) in(65′) should be determined from comparison with the experiment.

Expression (62) was obtained in the case of an isotropic effective mass of current carriers. Actually it has a more general character and does not change when anisotropy of effective masses is taken into account. For example, in layered HTSC materials the kinetic energy of current carriers in Hamiltonian (13) should be replaced by the expression:

$$T_a = \frac{1}{2m_{||}^*}\left(\hat{\boldsymbol{p}}_{1||} - \frac{e}{c}\boldsymbol{A}_1\right)^2 + \frac{1}{2m_{||}^*}\left(\hat{\boldsymbol{p}}_{2||} - \frac{e}{c}\boldsymbol{A}_2\right)^2 + \frac{1}{2m_{\perp}^*}\left(\hat{\boldsymbol{p}}_{1\perp} - \frac{e}{c}\boldsymbol{A}_{1z}\right)^2 + \frac{1}{2m_{\perp}^*}\left(\hat{\boldsymbol{p}}_{2\perp} - \frac{e}{c}\boldsymbol{A}_{1z}\right)^2, \tag{66}$$

where $\hat{p}_{1,2||}, \vec{A}_{1,2||}$ are the operators of the momentum and vector-potential in the planes of the layers (ab-planes); $\hat{p}_{1,2\perp},\ \vec{A}_{1,2\perp}$ are relevant values in the direction perpendicular to the planes (along the c-axis); $m^*_{||},\ m^*_{\perp}$ are effective masses in the planes and in the perpendicular direction.
As a result of the transformation:

$$\tilde{x} = x, \quad \tilde{y} = y, \quad \tilde{z} = \gamma z \tag{67}$$

$$\tilde{A}_{\tilde{x}} = A_x, \quad \tilde{A}_{\tilde{y}} = A_y, \quad \tilde{A}_{\tilde{z}} = \gamma^{-1} A_z,$$

$$\tilde{p}_{\tilde{x}} = p_x, \quad \tilde{p}_{\tilde{y}} = p_y, \quad \tilde{p}_{\tilde{z}} = \gamma^{-1} p_z,$$

where $(\gamma^*)^2 = m^*_{\perp}/m^*_{||}$, $\gamma^*$ is an anisotropy parameter, the kinetic energy $\tilde{T}_a$ turns out to be isotropic. Hence, $\tilde{\vec{\mathrm{P}}}_R + (2e/c)\tilde{\vec{A}}_{\tilde{R}} = 0$ . Then it follows from (67) that relation (62) is valid in the anisotropic case too. It follows that:

$$\boldsymbol{P}_{R||} = -\frac{2e}{c}\boldsymbol{A}_{R||}, \ \boldsymbol{P}_{R\perp} = -\frac{2e}{c}\boldsymbol{A}_{\perp},$$

$$\boldsymbol{j}_{||} = 2en_0\boldsymbol{P}_{R||}/M^*_{e||}, \qquad \boldsymbol{j}_{\perp} = 2en_0\boldsymbol{P}_{R\perp}/M^*_{e\perp}. \tag{68}$$

A magnetic field directed perpendicularly to the plane of the layers will induce currents flowing in the plane of the layers. When penetrating into the sample, this field will attenuate along the plane of the layers. For the magnetic field perpendicular to the plane of the layers ($H_\perp$), we denote the London penetration depth by $\lambda_\perp$ , and for the magnetic field parallel to the plane of the layers ($H_{||}$) – by $\lambda_{||}$.

This implies the expressions for the London depths of the magnetic field penetration into a sample:

$$\lambda_{\perp} = \left(\frac{M^*_{e\perp}c^2}{16\pi e^2 n_0}\right)^{1/2}, \qquad \lambda_{||} = \left(\frac{M^*_{e||}c^2}{16\pi e^2 n_0}\right)^{1/2}. \tag{69}$$

For $\lambda_{||}$ and $\lambda_{\perp}$ the denotations $\lambda_{ab}$ and $\lambda_c$ are often used. It follows from (69) that :

$$\frac{\lambda_{\perp}}{\lambda_{||}} = \left(\frac{M^*_{e\perp}}{M^*_{e||}}\right)^{1/2} = \gamma^*. \tag{70}$$

It also follows from (69) that the London penetration depth depends on the temperature:

$$\lambda^2(0)/\lambda^2(T) = n_0(T)/n_0(0). \tag{71}$$

In particular, for $\omega$=0, with the use of (54) we get: $\lambda(T)=\lambda(0)\ \left(1-(T/T_C)^{3/2}\right)^{-1/2}$. This dependence is compared with other approaches in Section 10.

It is usually believed that a Bose system becomes superconducting due to an interaction between the particles. The occurence of a gap in the spectrum of TI bipolarons can lead to their condensation even when the particles do not interact and the Landau superfluidity criterion

$$v = \hbar\omega_0/\mathrm{P} \tag{72}$$

(where Ƥ is a specific momentum of the bipolaron condensate) can be fulfilled even for non-interacting particles. From condition (72) we can derive the expression for the maximum value of the current density $j_{\max} = env_{\max}$:

$$j_{\max} = en_0\sqrt{\frac{\hbar\omega_0}{M_e^*}}. \qquad (73)$$

In conclusion it should be noted that all of the aforesaid refers to local electrodynamics. Accordingly, expressions obtained for λ are valid only on condition that λ $>> \xi$ , where $\xi$ is a correlation length which determines the characteristic size of a pair, i.e. the characteristic scale of changes in the wave function $\psi$(r) in (21). As a rule, this condition is fulfilled in HTSC. In ordinary superconductors the reverse inequality is fulfilled. A non-local generalization of superconductor electrodynamics was performed by Pippard [130]. It implies that the relation between $\vec{j}_S$ and $\vec{A}$ in expression (64) can be written in the form:

$$\boldsymbol{j}_S = \int Q(r - r')\boldsymbol{A}(r')d\boldsymbol{r}', \qquad (74)$$

where $Q$ is a certain operator whose radius of action is usually believed to be equal to $\xi$ . In the limit $\xi$ >>λ this leads to an increase in the absolute value of the depth of the magnetic field penetration into a superconductor which becomes equal to $(\lambda^2\xi)^{1/3}$ [130].

**9. Thermodynamic properties of a TI bipolaron gas in a magnetic field**.

The fact that Bose condensation of an ideal Bose gas in a magnetic field is impossible [131] does not mean that BEC mechanism cannot be used to describe superconductivity in a magnetic field. This follows from the fact that a magnetic field in a superconductor is identically zero. At the same time, abstracting ourselves from SC problem, there are no obstacles to consider a Bose gas to be placed in a magnetic field. Of interest is to investigate this problem in respect to a TI bipolaron gas.

First, it should be noted, that from expression for $\tilde{\omega}_H$, given by (53), it follows that for $\omega_0 = 0$, Bose condensation of TI bipolarons turns out to be impossible if $H \neq 0$. For an ordinary ideal charged Bose gas, this conclusion was first made in [131]. In view of the fact that in the spectrum of TI bipolarons there is a gap between the ground and excited states (Section 5), for a TI bipolaron gas this conclusion is invalid at $\omega_0 \neq 0$.

From the expression for $\tilde{\omega}_H$, (53) it follows that there is a maximum value of the magnetic field $H_{max}$equal to:

$$H_{max}^2 = \frac{2\omega_0\hbar^2 M_e}{\eta^2}. \qquad (75)$$

For $H > H_{max}$ , a homogeneous superconducting state is impossible. As suggested by (16), the quantity $\eta$ consists from two parts $\eta = \eta' + \eta''$. The value of $\eta'$ is determined by the integral involved into the expression for $\tilde{\omega}_k$ (28). Therefore $\eta'$ depends on the shape of a sample surface. The value of $\eta''$ is determined by the sum involved into the expression for $\tilde{\omega}_k$ (37) and depends on the shape of a sample surface only slightly. Hence, the value of $\eta$ can change as the shape of a sample surface changes thus leading to a change in $H_{max}$. With the use of (75), $\tilde{\omega}_H$ (53) will be written as:

$$\tilde{\omega}_H = \tilde{\omega}(1 - H^2/H_{max}^2). \quad (76)$$

For a given temperature T, let us write $H_{cr}$(T) for the value of the magnetic field for which the superconductivity disappears. This value of the field, according to (76), corresponds to $\tilde{\omega}_{H_{cr}}$:

$$\tilde{\omega}_{H_{cr}}(T) = \tilde{\omega}(1 - H_{cr}^2(T)/H_{max}^2). \quad (77)$$

The temperature dependence of $\tilde{\omega}_{H_{cr}}(T)$ can be found from condition (53):

$$C_{bp} = \tilde{T}^{\frac{3}{2}} F_{\frac{3}{2}}\left(\tilde{\omega}_{H_{cr}}(\tilde{T})/\tilde{T}\right).$$

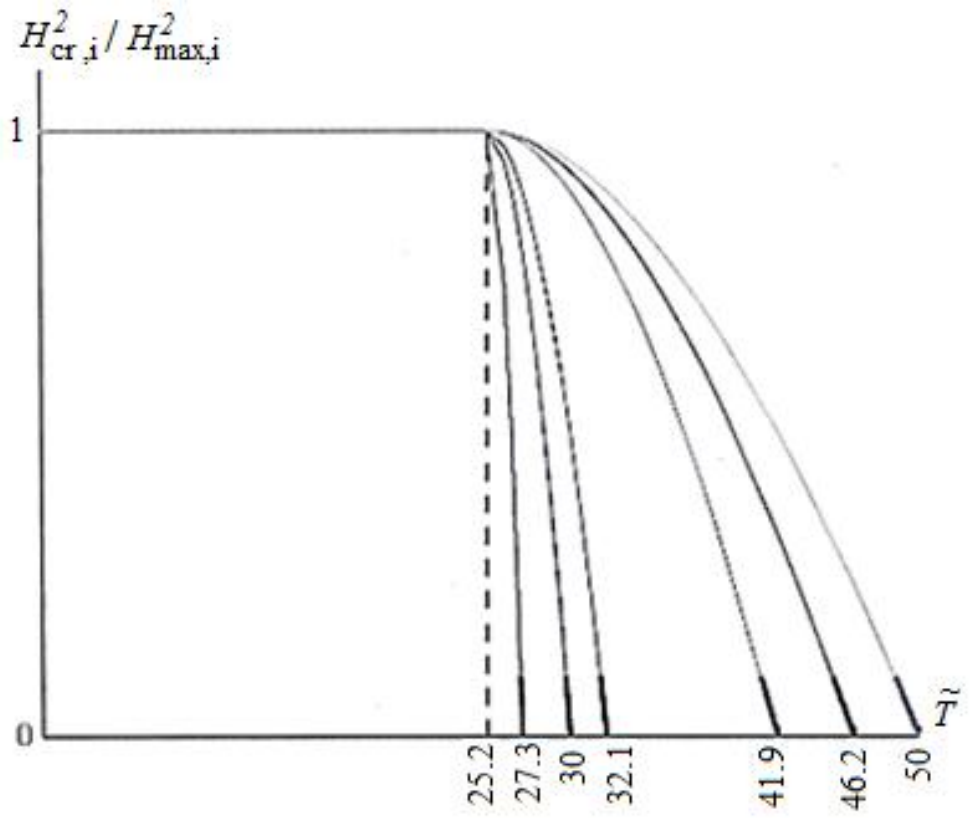

Fig.6 Temperature dependence $H_{cr,i}^2/H_{max,I}^2$ on the interval $[0; T_{c,i}]$ for the parameter values of $\tilde{\omega}_i$, given in Fig 2

It has the shape shown in Fig.2.

With the use of (75) and the temperature dependence given in Fig.2 we can find the temperature dependence of $H_{cr}(\tilde{T})$:

$$\frac{H_{cr}^2(\tilde{T})}{H_{max}^2} = 1 - \omega_{H_{cr}}(\tilde{T})/\tilde{\omega}. \quad (78)$$

For $\tilde{T} \leq \tilde{T}_{ci}$ these dependencies are given in Fig. 6.

According to Fig. 6, $H_{cr}(\tilde{T})$ reaches its maximum at a finite temperature $\tilde{T}_c(\tilde{\omega} = 0) \leq \tilde{T}_c(\omega_{0i})$. Fig.6 suggests that at a temperature below $\tilde{T}_c(\tilde{\omega} = 0) = 25.24$ a further decrease of the temperature no longer changes of the critical field $H_{cr}(\tilde{T})$ irrespective of the gap value $\tilde{\omega}$ .

Let us introduce a concept of a transition temperature $T_c(H)$ in a magnetic field $H$. Fig.7 shows the dependencies $T_c(H)$ resulting from Fig.6 and determined by the relations:

$$C_{bp} = \tilde{T}_{c,i}^{3/2}(H) F_{3/2}\left(\tilde{\omega}_{H,i}/\tilde{T}_{c,i}(H)\right), \;\; \tilde{\omega}_{H,i} = \tilde{\omega}_{H=0,i}\left[1 - H^2/H_{max,i}^2\right].$$

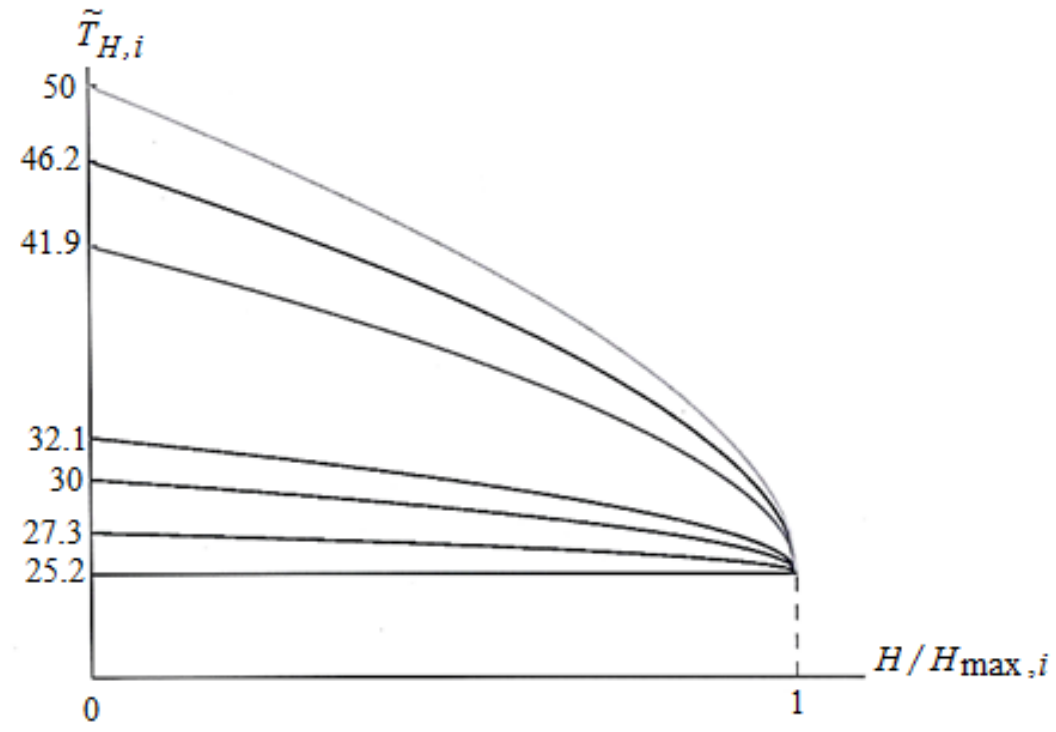

Fig.7. Dependence of the critical transition temperature $\tilde{T}_{H,i}$ on the magnetic field $H$ for $\widetilde{\omega}_i$parameter values, used in Fig.2.

Fig.7 suggests that the critical temperature of the transition $\tilde{T}_c(H)$ changes in a stepwise fashion as the magnetic field reaches the value $H_{max,i}$.

To solve the problem of the type of a phase transition in a magnetic field we will proceed from the well-known expression which relates the free energies in the superconducting and normal states:

$$F_s + \frac{H^2}{8\pi} = F_{N,} \tag{79}$$

where $F_s$ and $F_N$ are free energies of the unit volume of superconducting and normal states, respectively.

$$F_s = \frac{N}{V} E_{bp}(H = 0) - \frac{2}{3} \Delta E(\omega_{H=0}) \frac{N}{V},$$

$$F_N = \frac{N}{V} E_{bp}(H) - \frac{2}{3} \Delta E(\omega_H) \frac{N}{V},$$

where $\Delta E = E - E_{bp},\ E = \omega^* \tilde{E},$ where $\tilde{E}$ is determined by formula (58). Differentiating (79) with respect to temperature and taking into account that $S = -\partial F/\partial T$, we express the transition heat $q$ as:

$$q = T(S_N - S_S) = -T\partial(F_N - F_S)/\partial T = -T \frac{H_{cr}}{4\pi} \frac{\partial H_{cr}}{\partial T}. \tag{80}$$

Accordingly, the entropy difference $S_S - S_N$ is:

$$S_S - S_N = \frac{H_{cr}}{4\pi} \left( \frac{\partial H_{cr}}{\partial T} \right) = \frac{H_{max}^2}{8\pi\omega^*} (\tilde{S}_S - \tilde{S}_N). \tag{81}$$

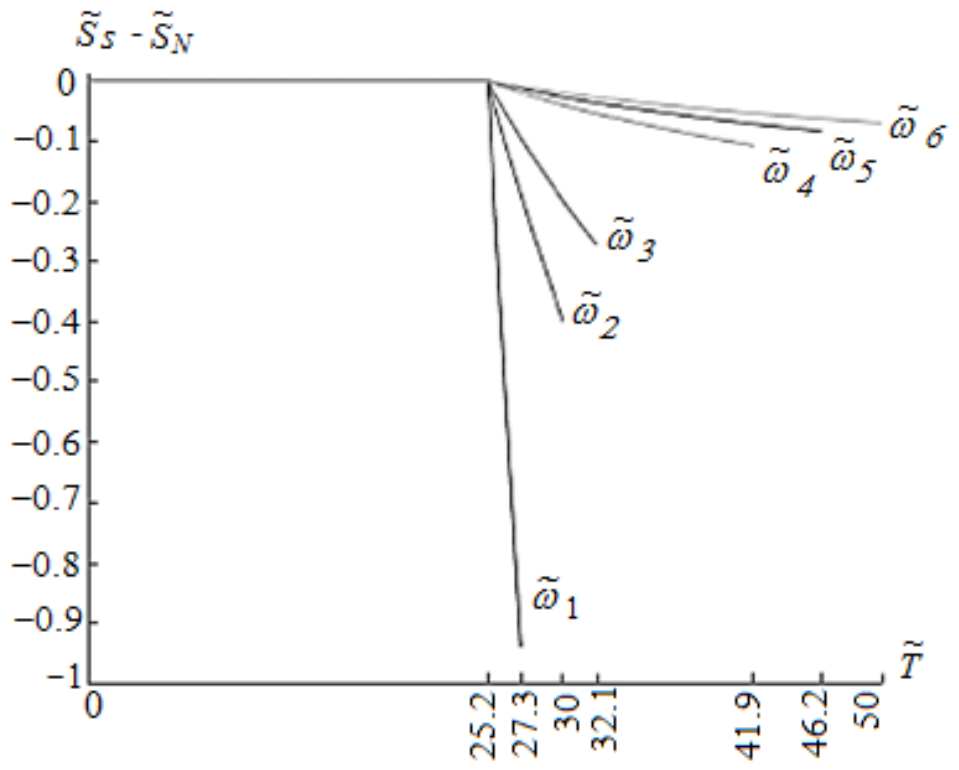

Fig.8. Temperature dependencies of differences between the entropies in the superconducting and normal states for $\tilde{\omega}_i$, parameters used in Fig.6,7.

Fig.8 illustrates the temperature dependence of differences between the entropies in the superconducting and normal states (81) for different values of critical temperatures ($\tilde{T}_i$) given in Fig.2. The differences presented can seem strange in, at least, two respects.

1. In the BCS and Landau theories at the critical point $T_c$ the entropy difference vanishes according to Rutgers formula. Entropy in Fig. 8 is a monotonous function $\tilde{T}$ and does nor vanish at $T$=$T_c$.

2. The absolute value of the difference $|\tilde{S}_S - \tilde{S}_N|$, when approaching the limit point $\tilde{T}_c = 25.2$, which corresponds to $\tilde{\omega} = 0$, would seem to increase rather than to decrease vanishing for $\tilde{\omega} = 0$.

Regarding point 2, this is indeed the case for $|\tilde{S}_S - \tilde{S}_N|$, since the value of the maximum field $H_{max}$ and, accordingly, the multiplier $H_{max}/8\pi$, which relates the quantities $S_S - S_N$ and $\tilde{S}_S - \tilde{S}_N$ vanishes for $\tilde{\omega} = 0$.

Regarding point 1, as it will be shown below, Rutgers formula turns out to be inapplicable to a TI bipolaron Bose condensate. Table 2 lists the values of $\tilde{S}_S - \tilde{S}_N$ for critical temperatures corresponding to different values of $\tilde{\omega}_{H_{cr,i}}$.

The results obtained lead to some fundamental consequences:

1.The curve of $H_{cr}(T)$ dependence (Fig.6) has a zero derivative $dH_{cr}(T)/dT = 0$ at $T$=0. This result is in accordance with the Nernst theorem which implies that entropy determined by (80) is equal to zero at $T$=0.

2. According to Fig.6, $H_{cr}(T)$ monotonously decreases as $T$ grows for $T > T_c(\tilde{\omega} = 0)$, and does not change for $T \leq T_c(\tilde{\omega} = 0)$. Hence $\partial H_{cr}(T)/\partial T < 0$ for $T > T_c(\tilde{\omega} = 0)$, . Therefore on the temperature interval $[T_c(\tilde{\omega} = 0), T_c(\tilde{\omega})]$ $S_S < S_N$, and on the interval $[0, T_c(\tilde{\omega} = 0)]$ $S_S = S_N$.

This has some important implications:

1. Transition on the interval $[0, T_c(\tilde{\omega} = 0)]$ occurs without absorption or liberation of latent heat since in this case $S_S = S_N$. That is in the experiment it will be seen as a second order phase transition. Actually, on the interval $[0, T_c(\tilde{\omega} = 0)]$the phase transition into the superconducting state is an infinite-order phase transition since in this region any-order derivatives of the energy difference $F_S - F_N$ vanish according to (79) and Fig.6.

2. In a magnetic field on the interval $[T_c(\tilde{\omega} = 0),\ T_c(\tilde{\omega})]$ which corresponds to $S_S < S_N$, a transition from a superconducting to a normal state occurs with absorption of latent heat. On the contrary, in passing on from a normal to a superconducting state latent heat releases. The phase transition on the interval $[0, T_c(\tilde{\omega} = 0)]$is not attended by a release or absorption of latent heat, being an infinite-order phase transition.

With regard to the fact that the heat capacity of a substance is determined by the formula $C=T(\partial S/\partial T)$, a difference in the specific heat capacities in a superconducting and normal states, according to (81), is written as:

$$C_S - C_N = \frac{T}{4\pi}\left[\left(\frac{\partial H_{cr}}{\partial T}\right)^2 + H_{cr}\frac{\partial^2 H_{cr}}{\partial T^2}\right]. \quad (82)$$

The well-known Rutgers formula can be obtained from this expression if we put in (82) the critical field $H_{cr}(T_c) = 0$ for $T = T_c$ and leave in the bracket in the right-hand side of (82) only the first term:

$$(C_s - C_N)_R = \frac{T_c}{4\pi}\left(\frac{\partial H_{cr}}{\partial T}\right)^2_{T_c}.$$

It is easy to see that at the point $T = T_c$ the value of $\omega_{H_{cr}}$ determined by Fig.2 for all the temperature values has a finite derivative with respect to $T$ and therefore, according to (78), an infinite derivative $\partial H_{cr}/\partial T$ for $T = T_c$ . Hence, the second term in the square bracket in (82) turns to -∞, leaving this bracket a finite value. As a result, the difference in the heat capacities in our model of Bose gas is properly expressed as:

$$C_S - C_N = \frac{T}{8\pi}\frac{\partial^2}{\partial T^2}H_{cr}^2(T) = \frac{H_{max}^2}{8\pi\omega^*}\left(\tilde{C}_S - \tilde{C}_N\right), \quad (83)$$

$$\tilde{C}_S - \tilde{C}_N = \tilde{T}\frac{\partial^2}{\partial \tilde{T}^2}\left(H_{cr}^2(\tilde{T})/H_{max}^2\right).$$

Table 2 lists the values of the quantity $\tilde{C}_S - \tilde{C}_N$ for the values of the critical temperatures corresponding to different values of $\tilde{\omega}_{H_{cr,i}}$. It should be noted that according to the results obtained, the maximum of the heat capacity jump occurs at a zero magnetic field and decreases as the magnetic field grows, vanishing for H = $H_{cr}$ which is fully consistent with the experimental data (Section 11). Comparison of the heat capacity jumps shown in Fig.5 with expression (83) enables us to calculate the value of $H_{max}$. The values of $H_{max}$ obtained for different values of $\omega_i$ are listed in Table 2. These values unambiguously determine the values of constant η in formulae (38'), (38'').

Table 2. The values of $H_{max}$, entropy differences $\tilde{S}_S - \tilde{S}_N$ and heat capacities $\tilde{C}_S - \tilde{C}_V$ in a superconducting and normal states determined by relations (81) and (83) are presented for the transition temperatures $\tilde{T}_{C_i}$, for the same values of $\tilde{\omega}_{H_{cr,i}}$, as in Fig.2.

| $i$ | $\tilde{\omega}_{H_{cr,i}}$ | $\tilde{T}_{C_i}$ | $\tilde{S}_S - \tilde{S}_N$ | $\tilde{C}_S - \tilde{C}_V$ | $H_{max} \cdot 10^{-3}$, Oe |
|---|---|---|---|---|---|
| 0 | 0 | 25.2 | 0 | 0 | 0 |
| 1 | 0.2 | 27.3 | -0.94 | -11.54 | 2.27 |
| 2 | 1 | 30 | -0.4 | -2.18 | 7.8 |
| 3 | 2 | 32.1 | -0.27 | -1.05 | 13.3 |
| 4 | 10 | 41.9 | -0.1 | -0.19 | 47.1 |
| 5 | 15 | 46.2 | -0.08 | -0.12 | 64.9 |
| 6 | 20 | 50 | -0.07 | -0.09 | 81.5 |

It follows from what has been said that Ginzburg–Landau temperature expansion for a critical field near a critical temperature $T_c$ is inapplicable to a TI bipolaron Bose condensate. Since the temperature dependence $H_{cr}(T)$ determines the temperature dependencies of all the thermodynamic quantities, this conclusion is valid for all such quantities. As noted in the Introduction, this conclusion stems from the fact that the BCS theory, being non-analytical with respect to the coupling constant under no conditions passes on the bipolaron condensate theory.
Above we dealt with the isotropic case. In the anisotropic case it follows from formulae (38'), (38'') that:

$$H_{max}^2 = H_{max\perp}^2 = \frac{2\omega_0 M_\perp \hbar^2}{\eta^2}, \qquad \boldsymbol{B}||\boldsymbol{c}, \tag{84}$$

i.e. in the case when a magnetic field is directed perpendicularly to the plane of the layers and:

$$H_{max}^2 = H_{max||}^2 = \frac{2\omega_0 M_{||} \hbar^2}{\eta^2}, \qquad \boldsymbol{B} \perp \boldsymbol{c}, \tag{85}$$

in the case when a magnetic field lies in the plane of the layers. From (84), (85) it follows that:

$$\frac{H_{max\perp}^2}{H_{max||}^2} = \sqrt{\frac{M_\perp}{M_{||}}} = \gamma. \tag{86}$$

With the use of (78), (85), (86) we get for the critical field $H_{cr}(\tilde{T})$ in the directions perpendicular and parallel to the plane of the layers:

$$H_{cr||,\perp}(\tilde{T}) = H_{max||,\perp}\sqrt{1 - \omega_{H_{cr}}(\tilde{T})/\tilde{\omega}}. \tag{87}$$

It follows from (69) that relations $H_{cr||}(\tilde{T})/H_{cr\perp}(\tilde{T})$ are independent of temperature. The dependencies obtained are compared with the experiment in Section 11.

## 10. Scaling relations.

Scaling relations play an important role in the superconductivity theory by promoting a search for new high-temperature superconductors with record parameters. Such relations can be a generalization of a lot of experiments having no reliable theoretical justification, or can be deduced from less-than-reliable theoretical construct, though being experimentally confirmed in the future. An example is provided by Uemura law considered in the next section.

The theory presented enables one to give a natural explanation to some important scaling relations. In particular, in this section we deduce Alexandrov formula [132]-[133] and Homes' scaling law.

**Alexandrov formula.** As it was noted above (Section 5), in an anisotropic case formula (53) takes on the form:

$$\tilde{T}_c = F_{3/2}^{-2/3}(\tilde{\omega}/\tilde{T}_c)\left(\frac{n_{bp}}{M_{||}}\right)\frac{2\pi\hbar^2}{M_\perp^{1/3}\omega^*}. \tag{88}$$

It is convenient to pass on from the quantities which are difficult to measure in an experiment $n_{bp}, M_{||}, M_{\perp}$ to those which can be measured experimentally:

$$\lambda_{ab} = \left[\frac{M_{||}}{16\pi n_{bp} e^2}\right]^{1/2}, \lambda_c = \left[\frac{M_{\perp}}{16\pi n_{bp} e^2}\right]^{1/2}, R_H = \frac{1}{2en_{bp}}, \qquad (89)$$

where $\lambda_{ab} = \lambda_{||,}$ $\lambda_c = \lambda_{\perp}$are London depths of penetration into the planes of the layers and in the perpendicular direction, accordingly; $R_H$ is Hall constant. In expressions (89) the light velocity is assumed to be equal to one: $c$=1. With the use of relations (89) and (88) we get:

$$k_B T_c = \frac{2^{1/3}}{8} F_{3/2}^{-2/3}\left(\tilde{\omega}/\tilde{T}_c\right) \frac{\hbar^2}{e^2} \left(\frac{eR_H}{\lambda_{ab}^4 \lambda_c^2}\right)^{1/3}. \qquad (90)$$

Here the value of $eR_H$ is measured in cm$^3$, $\lambda_{ab}$, $\lambda_c$ - in cm, $T_c$ - in K.

Taking into account that in most HTSC $\tilde{\omega} \approx \tilde{T}_c$ and function $F_{3/2}\left(\tilde{\omega}/\tilde{T}\right)$ varies in the vicinity of $\tilde{\omega} \approx \tilde{T}_c$ only slightly, with the use of the value $F_{\frac{3}{2}}(1) = 0.428$ we derive from (78) that $T_c$ is equal to:

$$T_c \cong 8.7 \left(\frac{eR_H}{\lambda_{ab}^4 \lambda_c^2}\right)^{\frac{1}{3}}. \qquad (91)$$

Formula (91) differs from Alexandrov formula [132]-[133] only in a numerical coefficient which is equal to 1,64 in [132]-[133]. As it is shown in [132]-[133], formula (91) properly describes, almost without exception, a relation between the parameters for all known HTSC materials. It follows from (88) that Uemura relation [134], [135] is a particular case of formula (91).

In an isotropic case, formulae (90), (91) also yield a well-known law of a linear dependence of $T_c$ on the inverse value of the squared London penetration depth.

**Homes' law.** Homes' law claims that in the case of superconducting materials scaling relation holds [136], [137]:

$$\rho_S = C\sigma_{DC}(T_C)T_C, \qquad (92)$$

where $\rho_S$ is the density of the superfluid component for $T$=0, $\sigma_{DC}(T_C)$ is the direct current conductivity for $T$=$T_c$, $C$ is a constant equal to ~35 cm$^{-2}$ for ordinary superconductors and HTSC for a current running in the plane of the layers.

The quantity $\rho_S$ in (92) is related to plasma frequency $\omega_P$ as $\rho_S = \omega_p^2$ [138]. ($\omega_p = \sqrt{4\pi n_S e_S^2/m_S^*}$, where $n_S$ is a concentration of superconducting current carriers; $m_S^*$, $e_S$ - are a mass and a charge of superconducting current carriers). With the use of this relation, relation $\sigma_{DC} = e_n^2 n_n \tau/m_n^*$ (where $n_n$ is a concentration of current carriers for $T = T_c$, $m_n^*$, $e_n$, is a mass and charge of current carriers for $T$=$T_c$ ), and relation $\tau \sim \hbar/T_c$ (where $\tau$ is a minimum Plank time of electron scattering at a critical point) [138], on the assumption that $e_S = e_n$, $m_S = m_n$, we get from (92):

$$n_S(0) \cong n_n(T_c). \qquad (93)$$

In our scenario of Bose condensation of TI bipolarons, Homes' law in the form of (97) becomes almost obvious. Indeed, for $T=T_c$ , TI bipolarons are stable (they decay at a temperature equal to the pseudogap energy which far exceeds $T_c$ ). Their concentration at $T=T_c$ is equal to $n_n$ and, therefore, at $T=T_c$ these bipolarons start forming a condensate whose concentration $n_S(T)$ reaches maximum $n_S(0)=n_n(T_c)$ at $T=0$ (i.e. when bipolarons become fully condensed) which corresponds to relation (93). It should be noted that in the framework of the BCS theory Homes' law cannot be explained.

## 11. Comparison with the experiment.

Success of the BCS theory is concerned with successful explanation of some experiments in ordinary metal superconductors where EPI is not strong. It is arguable that EPI in high-temperature ceramic SC is rather strong [139]-[141] and the BCS theory is hardly applicable to them. In this case it may be worthwhile to use the description of HTSC properties on the basis of bipolaron theory. As is known, Eliashberg theory which was developed to describe SC with strong EPI [25] is inapplicable to describe polaron states [142]-[143]. Let us list some experiments on HTSC which are in agreement with the TI bipolaron theory.

According to the main currently available SC theories (BCS, RVB, t-J theories [16], [106]-[107]), at low temperatures all the current carriers should be paired (i.e. the superconducting electron density coincides with the superfluid one). In recent experiments on overdoped SC [43] it was shown that this is not the case – only a small portion of current carriers were paired. The analysis of this situation performed in [144] demonstrates that the results obtained in [43] do not fit in the available theoretical constructions. The TI bipolaron theory of SC presented above gives an answer to the question of paper [144] – where most of the electrons in the studied SC disappeared? The answer is that only a small portion of electrons $n_{bp}$: $n_{bp}\approx n|E_{bp}|/E_F\ll n$, occurring near the Fermi surface are paired and determine the surface properties of HTSC materials.

Actually, however, the theory of EPI developed in that work is applicable to underdoped SC and inapplicable to describe experiments with overdoped samples which were used in [43]. In particular, in underdoped samples, we cannot expect a linear dependence of the critical temperature on the density of SC electrons which was observed in [43]. This dependence should rather be expected to be nonlinear, as it follows from equation (53).

To describe the overdoped regime a theory [145] has recently been constructed on the basis of Fermi condensation described in [146]. It is a generalization of the BCS theory where it was shown that the number of SC current carriers is only a small portion of their total number which is in agreement with the results of [43].

Hence we can conclude that the results obtained in [43] are rather general and are valid for both underdoped and overdoped cases (see also [147]).

We can also expect that the temperature dependence of the resistance is linear for $T > T_c$ in the underdoped and overdoped cases since the number of bipolarons is small as compared to the total number of electrons, if EPI is dominant and a crystal is isotropic.

In contrast to paper [145] in recent work [148] it was shown that the linear dependence of $T_c$ on the number of Cooper pairs which was observed in [43] for overdoped $L_{a_{2-x}}S_{r_x}C_{u_2}O$ crystals can be explained in terms of the BCS on the basis of plasmon mechanism of SC. Nevertheless it seems that the special case considered in [145] cannot explain the general character of the results obtained in [43].

The problem of inability of the BCS and other theories to explain the results of [43] was also considered in recent work [149] where a simple model of a bipolaron SC is developed and the number of bipolaron current carriers is shown to be small as compared to the total number of electrons. The results obtained in [149] confirm the results of [109],[117],[118] that the portion of paired states is small in the low-temperature limit.

Important evidence in favor of bipolaron mechanism of SC is provided by experiments on measuring the noise of tunnel current in LSCO/LCO/LSCO heterostructures performed in [150].

According to these experiments, paired states of current carriers exist at $T>T_c$ too, i.e. they form before the formation of a superconducting phase. This crucially confirms the applicability of the bipolaron scenario to high-temperature oxides.

Fig. 4 illustrates typical dependencies of $E(\tilde{T})$. They suggest that at the transition point the energy is a continuous function $\tilde{T}$. This means that the transition *per se* proceeds without expending energy and the transition is the second order phase transition in full agreement with the experiment. At the same time the transition of Bose particles from the condensed state to the supracondensed one proceeds with consuming energy which is determined by quantity q (Section 3, Table 1) which determines latent transition heat of Bose gas, therefore the first order phase transition takes place.

Let us consider $Y\,B_{a_2}C_{u_3}O_7$ HTSC with the transition temperature 90÷93 K, the unit cell volume $1734\cdot 10^{-21}\text{cm}^3$, hole concentration $n \approx 10^{21}\text{cm}^{-3}$. According to estimates made in [151], the Fermi energy is equal to $E_F$ =0,37 eV. The concentration of TI bipolarons in$Y\,B_{a_2}C_{u_3}O_7$ can be found from equation (53):

$$\frac{n_{bp}}{n}C_{bp} = f_{\tilde{\omega}}(\tilde{T}_c)$$

with $c\,\tilde{T}_c = 1{,}6$. Table 1 lists the values of $n_{bp,i}$ for $\tilde{\omega}_i$ parameters presented in Section 6. Table 1 suggests that $n_{bp,i} \ll n$. Hence only a small portion of current carriers are in the bipolaron state. It follows that, in full agreement with the results of Section 6, the Coulomb interaction of bipolarons will be screened by unpaired electrons, which justifies the approximation of a non-interacting TI bipolaron gas considered.

According to this approach, for a SC to arise paired states should form. The condition of the formation of such states in the vicinity of the Fermi surface, according to [119] has the form: $E_{bp}<0$. Accordingly, the value of the pseudogap, according to the results of Section 6, will be:

$$\Delta_1 = \left|E_{bp} + u_0\right| . \qquad (94)$$

Naturally, this value is independent of the vector $\vec{k}$ but depends on the concentration of current carriers, i.e. the level of doping.

In the simplest version of the SC theory under consideration, the gap $\omega_0$ does not change in passing on from the condensed to the non-condensed state, i.e. in passing on from the superconducting to the nonsuperconducting state and, therefore, $\omega_0$ has also the meaning of a pseudogap:

$$\Delta_2 = \omega_0(\boldsymbol{k}), \qquad (95)$$

which depends on the wave vector $\vec{k}$.

Numerous discussions on the gap and pseudogap problem stem from the statement that the energy gap in HTSC is determined by the coupling energy of Cooper pairs which leads to insoluble contradictions (see reviews [152]-[156]).

Actually, the value of a SC gap $\Delta_2$ determined by (95), generally speaking, does not have anything to do with the energy of paired states which is determined by $E_{bp}$. According to [116], for small values of the EPI constant α and for large ones, the bipolaron energy $\left|E_{bp}\right| \sim \alpha^2\omega_0$, i.e. $\left|E_{bp}\right|$ does not depend on $\omega_0$.

For example, in the framework of the concept considered, it is clear why the pseudogap $\Delta_2$ has the same anisotropy as the SC gap – this is one and the same gap. It is also clear why the gap and the pseudogap depend on temperature only slightly. In particular, it becomes understandable why in the course of a SC transition a gap arises immediately and does not vanish at $T = Tc$ (this is not BCS behaviour). Much-debated question of what order parameter should be put into correspondence to the pseudogap phase (i.e. whether the pseudogap phase a special state of the matter [153]) seems to be senseless within the theory presented.

Presently there are a lot of methods for measuring a gap: angle resolved photo electron spectroscopy (ARPES), Raman (combination) spectroscopy, tunnel scanning spectroscopy, magnet neutron scattering, etc. According to [156], for the maximum value of the gap in $Y\,BCO$ (6.6) (in the antinodal direction in the ab-plane), it was obtained $\Delta_1/T_c \approx 16$. This yields $|E_{bp}|$ ≈80 meV.

Let us determine the characteristic energy of phonons responsible for the formation of TI bipolarons and superconducting properties of oxide ceramics, i.e. the value of a SC gap $\Delta_2$. To do so we compare the calculated jumps of the heat capacities with the experimental values.

A theoretically-calculated jump of the heat capacity (Fig.5, Section 6) coincides with the experimental values of jumps in $Y\,B_{a_2}C_{u_3}O_7$ [157] for $\tilde{\omega} = 1{,}5$, i.e. for $\omega$ = 7,5 meV. This corresponds to the TI bipolaron concentration $n_{bp} = 2{,}6 \cdot 10^{18}\text{cm}^{-3}$. Taking into account that $|E_{bp}| \approx 0.44\alpha^2\omega$ [44], $|E_{bp}|$= 80 meV, ω = 7,5 meV, the EPI constant α will be: α≈ 5, which is far beyond the limits of the BCS applicability.

As is known, in the BCS theory a jump of the heat capacity is equal to: $(C_S - C_n)/C_n = 1.43$, (where $C_S$is the heat capacity in the superconducting phase, and $C_n$is that in the normal one) and does not depend on the model Hamiltonian parameters. As it follows from numerical calculations presented in Fig. 5 and Table 1, as distinct from the BCS, the jump value depends on the phonon frequency. Hence, the approach presented predicts the existence of the isotopic effect for the heat capacity jump.

It should be noted that in calculations of the transition temperature it was believed that the effective mass $M_e$ in equation (53) is independent of the wave vector direction, i.e. an isotropic case was considered.

In an anisotropic case, choosing the main axes of vector $\vec{k}$ for the coordinate axes, we will get $\left(M_{ex}M_{ey}M_{ez}\right)^{1/3}$instead of the effective mass $M_e$. In layered HTSC materials the values of effective masses lying in the plane of the layers $M_{ex}, M_{ey}$ are close in value. Assuming that $M_e = M_{ex} = M_{ey} = M_{||}, M_{ez} = M_{\perp}$, we get instead of $C_{bp}$ determined by (53), the quantity $\tilde{C}_{bp} = C_{bp}/\gamma$, $\gamma^2 = M_{\perp}/M_{||}$ is the anisotropy parameter. Hence consideration of the anisotropy of effective masses gives for concentration $n_{bp}$ the value $\tilde{n}_{bp} = \gamma n_{bp}$. Therefore consideration of anisotropy can enlarge the estimate of the TI bipolaron concentration by an order of magnitude and greater. If for $YBa_2Cu_3O_7$ we take the estimate $(\gamma^*)^2 = 30$ [158], then for the TI bipolaron concentration we get $\tilde{n}_{bp} = 1.4 \cdot 10^{19}\text{cm}^{-3}$ , which leaves in place the main conclusion: in the case considered only a small portion of current carriers are in the TI bipolaron state. The situation can change if the anisotropy parameter is very large. For example, in layered HTSC Bi-Sr-Ca-Cu-O the anisotropy parameter is γ>100, therefore the concentration of TI bipolarons in these substances can be of the same order as the concentration of current carriers.

Another important conclusion suggested by consideration of anisotropy of effective masses is that the transition temperature $T_c$ depends not on $n_{bp}$ and $M_{||}$ individually, but on their relation which straightforwardly follows from (53). This phenomenon in known as Uemura law. In the previous section we discussed a more general relation, known as Alexandrov formula (for which Uemura law is a particular case).

Among the experiments involving an external magnetic field, of great importance are those on measuring the London penetration depth λ. In $YBa_2Cu_3O_7$ for λ at $T$=0 in paper [159] it was obtained that $\lambda_{ab} = 150 \div 300$ nm, $\lambda_c = 800$ nm. The same order of magnitude is obtained for these quantities in a number of works [160]-[163]. In paper [162] (see also references therein) it is shown that anisotropy of depths $\lambda_a$ and $\lambda_b$ in cuprate planes can account for 30% depending on the type of the crystal structure. If we take the values $\lambda_a = 150\, nm$ and $\lambda_c = 800\, nm$, obtained in most papers, then the anisotropy parameter, by (70), will be $\gamma^2 \approx 30$ . This value is usually used for $YBa_2Cu_3O_7$ crystals.

The temperature dependence $\lambda^2(0)/\lambda^2(T)$ was studied in many works (see [163] and references therein). Fig.9 compares different curves for $\lambda^2(0)/\lambda^2(T)$ . In paper [163] it was shown that in high-quality $YBa_2Cu_3O_7$ crystals the temperature dependence $\lambda^2(0)/\lambda^2(T)$ is well approximated by a simple dependence $1 - t^2, t = T/T_c$. Fig. 10 illustrates a comparison of the experimental dependence $\lambda^2(0)/\lambda^2(T)$ with the theoretical one:

$$\frac{\lambda^2(0)}{\lambda^2(T)} = 1 - \left(\frac{T}{T_c}\right)^{3/2} \frac{F_{3/2}(\omega/T)}{F_{3/2}(\omega/T_c)}, \tag{96}$$

which results from (58), (35), (34). Hence there is a good agreement between the experiment and the theory (96).

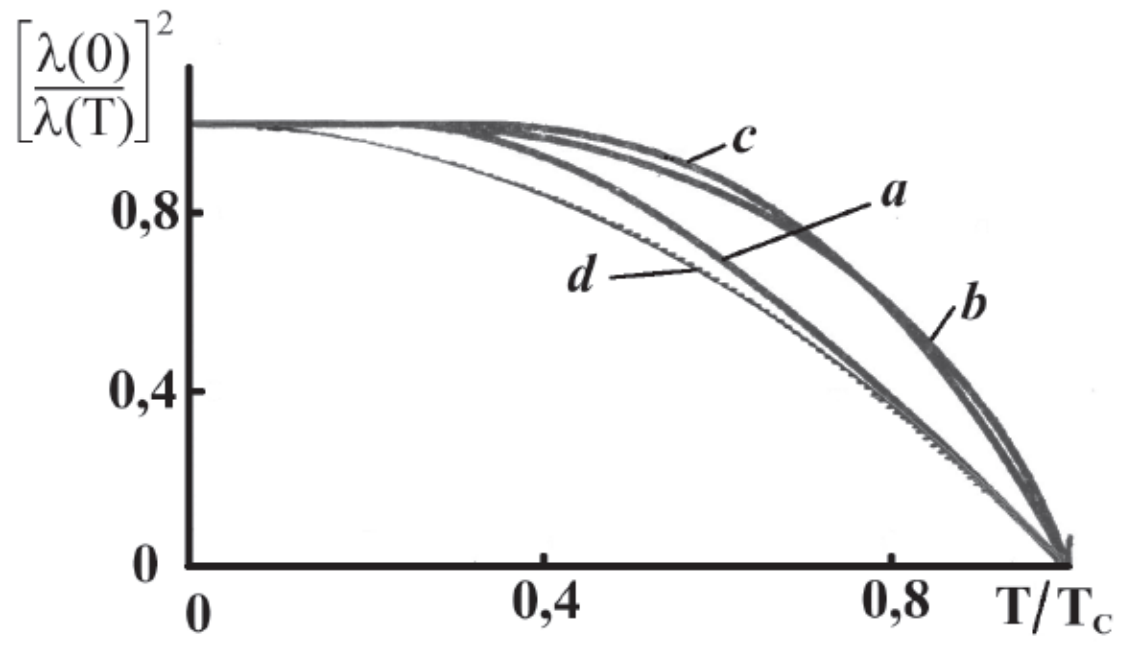

Fig. 9. Depth of a magnetic field penetration in the BCS theory (a – local approximation, b – non-local approximation); empirical rule $\lambda^{-2} \sim 1 - (T/T_c)^4$ (c) [164]; in $YBa_2Cu_3O_7$ [163].

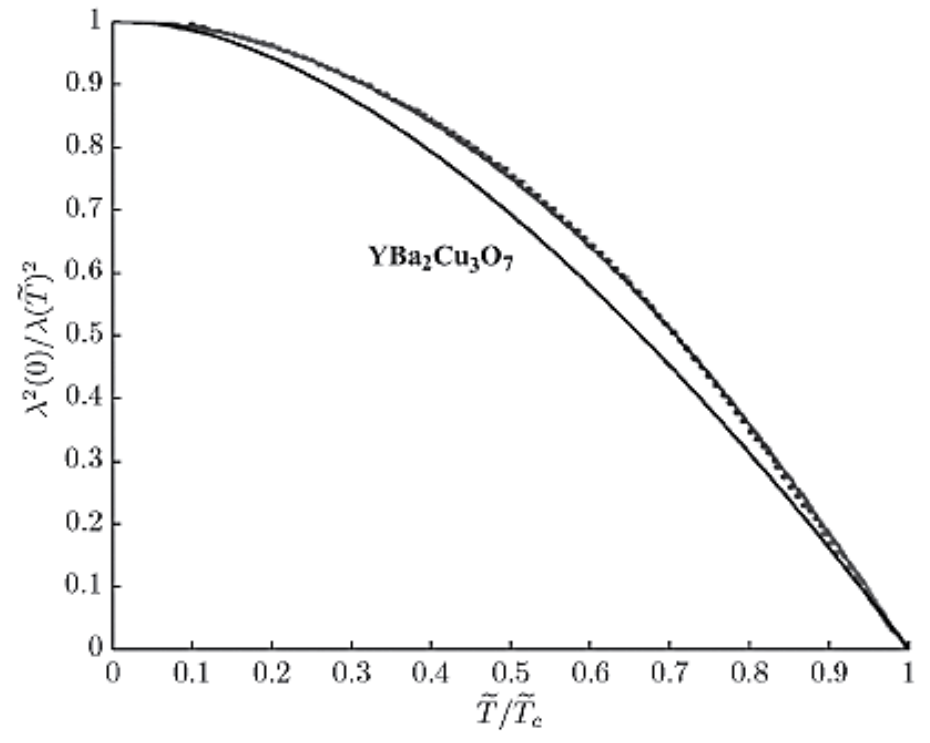

Fig.10. Comparison of the theoretical dependence $\lambda^2(0)/\lambda^2(\tilde{T})$ (solid curve), obtained in this paper with the experimental one [163] (dotted curve).

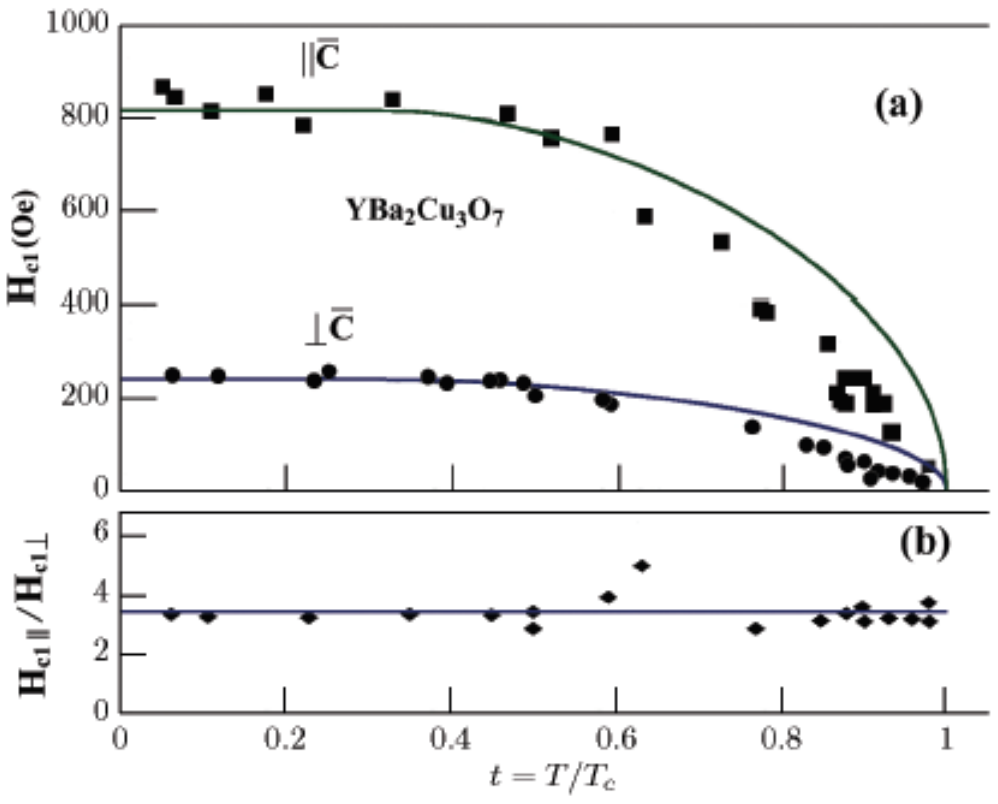

Fig.11. Comparison of calculated (continuous curves) and experimental values of $H_{c1}$ (squares, circles, rhombs [165] for the cases $||\vec{c}$ and $\perp \vec{c}$.

The theory developed enables one to compare the temperature dependence of the critical magnetic field in $Y\,Ba_2Cu_3O_7$with the experiment [165]. Since the theory constructed in Section 9 describes a homogeneous state of a TI bipolaron gas, the critical field being considered corresponds to the homogeneous Meissner phase. In paper [165] this field is denoted as $H_{c1}$, which relates to the denotations of Section 9 as: $H_{c1} = H_{cr}, H_{c1||} = H_{cr\perp}, H_{c1\perp} = H_{cr||}$. For comparison with the experiment, we use the parameter values earlier obtained for $Y\,Ba_2Cu_3O_7$: $\tilde{\omega} = 1.5,\ \tilde{T}_c = 1.6$. Fig.11 shows a comparison of the experimental dependencies of $H_{c1\perp}(T)$ and $H_{c1||}(T)$ [165] with the theoretical ones (87), where for $H_{\max||,\perp}(T)$ the experimental values: $H_{\max||} = 240$, $H_{\max\perp} = 816$ are taken. The results presented in Fig.11 confirm the conclusion (Section 9) that relations $H_{cr\perp}(T)/H_{cr||}(T)$ are independent of temperature.

It follows from (70), (84), (85) that:

$$(\gamma^*)^2 = \frac{M_\perp^*}{M_{||}^*} \propto \frac{\lambda_\perp^2}{\lambda_{||}^2};\ \ \frac{H_{max\perp}^2}{H_{\max||}^2} = \gamma^2 = 11.6. \qquad (97)$$

The choice of the value $\gamma^2 = 11.6$, determined by (97) for the anisotropy parameter differs from the value $(\gamma^*)^2 = 30$, used above. This difference is probably caused by a difference in the anisotropy of the polaron effective masses $M_{||,\perp}^*$ and electron band masses.

The presence of a gap $\omega_0$ in HTSC ceramics is proved by numerous spectroscopic experiments (ARPES) on angular dependence of $\omega_0\ on\ \vec{k}$ for small $|\vec{k}\ |$ [152]-[156]. The availability of d-symmetry in the angular dependence $\omega_0(\vec{k}\ )$ is probably concerned with the appearance of a pseudogap and transformation of Fermi system into the system of Fermi arcs possessing d-symmetry. In experiments on tunnel spectroscopy the quantity $\omega_0$ can manifest itself as an availability of a gap substructure superimposed on pseudogap $\Delta_1(\Delta_1 \gg \omega_0)$. This structure was frequently observed in optimally doped $Y\ Ba_2Cu_3O_7$ and $Bi_2Sr_2CaCu_2O_8$ (BCCO) in the range of 5÷10 meV [166]-[168] which coincides with the estimate of $\omega_0$presented above.

In a lot of experiments the dependence of the gap and pseudogap value on the level of doping x is measured. Even early experiments on magnetic susceptibility and Knight shift revealed the availability of the pseudogap which emerges for $T^* > T_c$. Numerous subsequent experiments revealed the peculiarities of the $T$ - $x$ phase diagram: $T^*$ increases and $T_c$ decreases as doping decreases [152]-[156]. As it is shown in [119], this behavior can be explained by peculiarities of the existence of bipolarons in a polaron gas.

It is noted in [119] that 1/8 anomaly (Fig.12) in HTSC systems [169] has probably general character.

The stability condition $E_{bp} < 0$ presented above means that the presence of Fermi gas radically changes the criterion of bipolaron stability which, in the absence of Fermi environment, takes on the form $E_{bp} < 2E_p$. This stabilization was first pointed out in papers [170] –[171]. This fact plays an important role in explaining concentration dependencies of $T_c$ on x. Most probably, in real HTSC materials the value of the HTSC constant has an intermediate value. Then in the range of small concentrations in the absence of Fermi environment, TI bipolarons are unstable with respect to their decay into individual polarons and SC at small x is impossible. It arises for finite x when there is a pronounced Fermi surface which stabilizes the formation of bipolarons. This corresponds to a lot of experiments on HTSC materials. A simple thermodynamic analysis demonstrates that at a finite temperature TI bipolarons are stable if $|E_{bp} - 2E_p| \gtrsim T$. Hence, the characteristic temperature $T^*$ corresponding to the pseudogap phase is equal to: $T^* \approx |E_{bp} - 2E_p|$

The transition to the pseudogap phase *per se* is concerned with the formation of TI bipolarons for $T < T^*$ and highly blurred with respect to temperature in full agreement with the experiment. It should be noted that $T^* << |E_{bp}|$ and approximately 1,5 ÷ 2 times exceeds $T_c$.

As doping increases at x > $x_{opt}$, where $x_{opt}$ is the value of optimal doping, SC passes on to overdoped regime when the number of bipolarons becomes so large that they start overlapping, i.e. a transition to the regime of BCS with small $T_c$ takes place.

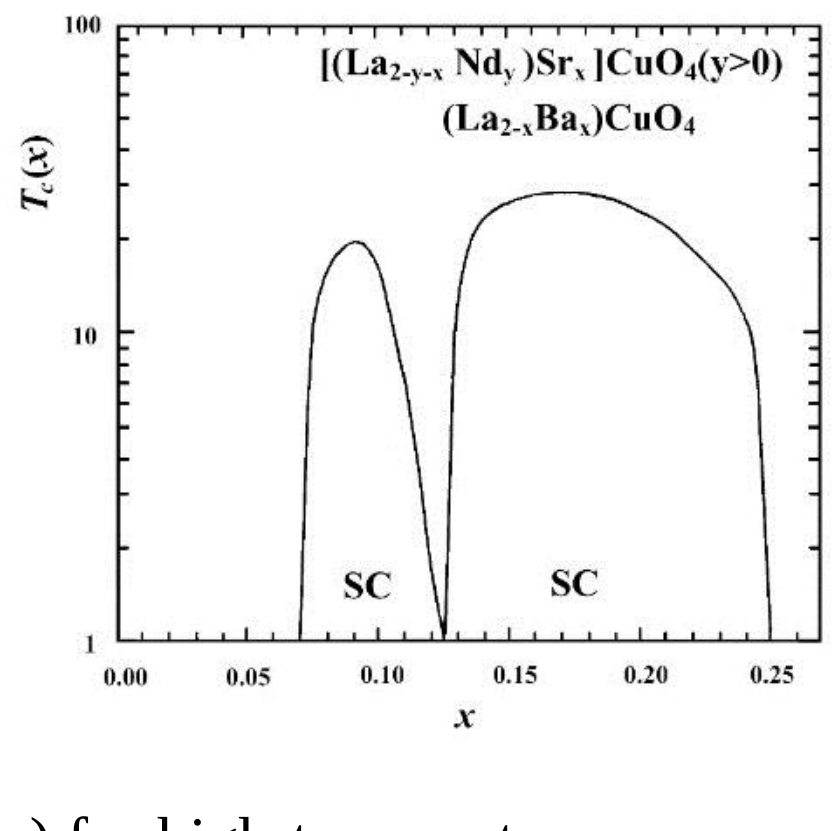

Fig. 12. Dependence of $T_c$(x) for high-temperature superconductors with 1/8 anomaly

In conclusion it should be noted that the long-term discussion of the nature of the gap and pseudogap in HTSC materials is largely related to the methodological problem of measurements when different measuring techniques actually measure not the same but different quantities. In the case under consideration ARPES measures $\omega_0(\vec{k})$, while tunnel spectroscopy $|E_{bp}|$. Below we consider these problems in greater detail.

## 12. Spectroscopic experiments.

As it is shown in the previous section, the theory developed is consistent with thermodynamic and magnet characteristics of HTSC materials. However, these facts are insufficient to judge unambiguously that the TI bipolaron theory of SC does not contradict other experimental facts.

Presently there are a lot of methods to study the properties of paired states and consequences of these states. The aim of this section is to analyze to what extent the data of modern spectroscopic methods such as scanning tunnel microscopy (STM), quasiparticle interference, angle resolved photo electron spectroscopy (ARPES) and Raman (combination) scattering are compatible with the ideas of the TI bipolaron mechanism of HTSC.

**Tunnel characteristics.** In the case of the TI bipolaron theory of SC tunnel characteristics have their peculiarities. As usual, in considering tunnel phenomena, for example, in considering a transition from a superconductor to an ordinary metal via a tunnel contact we will reckon the energy from the ground state of the SC. In the TI bipolaron theory of a SC, the ground state is the bipolaron state whose energy is below the Fermi level of this SC in the normal state by the value of the bipolaron energy $|E_B|$. Hence, as a result of a tunnel contact of a SC with an conventional metal the Fermi level of an conventional metal will coincide with the ground state energy of a SC. It follows that the one-particle current will have the usual form for such a contact (Fig. 13).

A peculiarity arises in considering a two-particle current. It is concerned with the fact that the spectrum of excited states of a TI bipolaron is separated from the ground state by the value of the phonon frequency $\omega_0$. For this reason, the volt-ampere characteristic of a two-particle current will have the form shown in Fig. 14, where $|E_B/2|$ is replaced by $\omega_0$. As a result the resulting volt-ampere characteristic will have the form of Fig. 14.

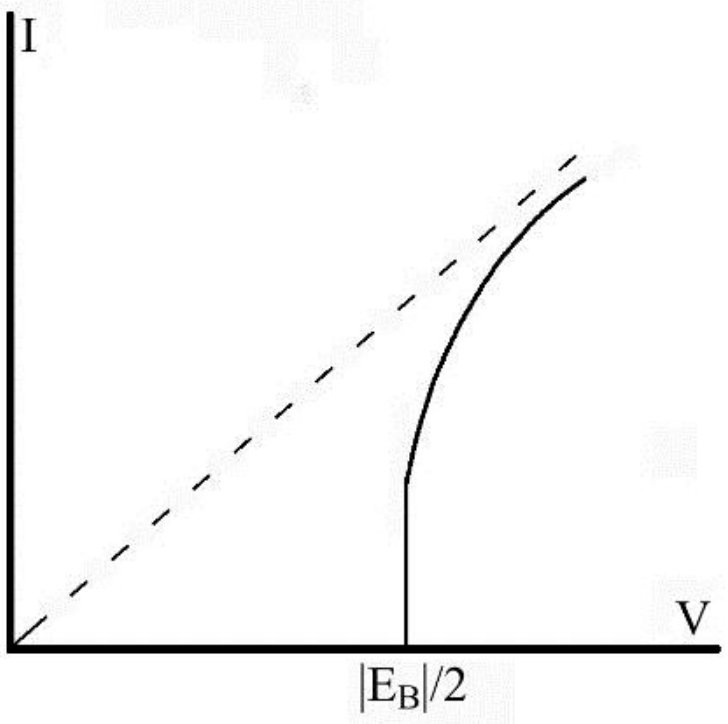

*Fig. 13.* Volt-ampere characteristic of a one-particle current

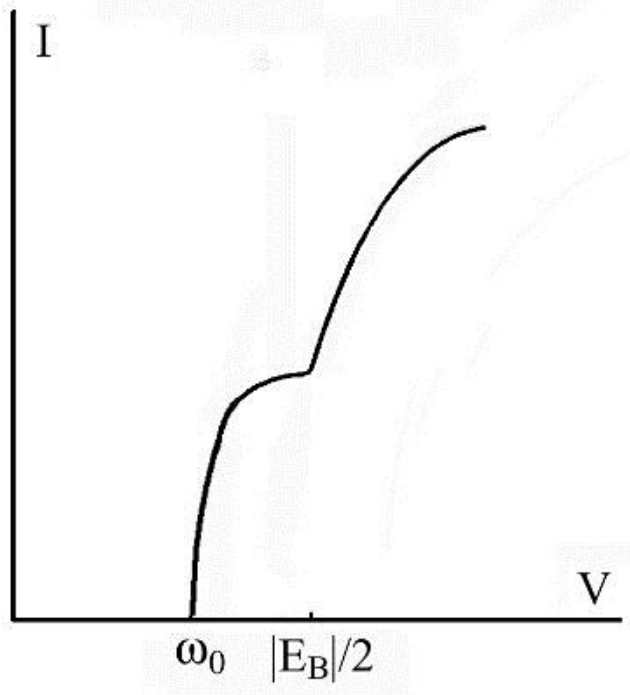

*Fig. 14.* Volt-ampere characteristic of a total current

The curve *I-V* is constructed for the case of $\omega_0 < |E_B|/2$. In the opposite case the quantities $\omega_0$ and $|E_B|/2$ should be inverted. The $\omega_0 < V < |E_B|/2$ segment of the *I-V* curve in Fig. 14, corresponds to a kink which is lacking in the BCS theory.

Spectrally, a kink corresponds to a transition of a one-particle electron spectrum with energy lying lower than $E_F$ by the value of $|E_B|/2$ to a two-particle TI bipolaron spectrum of excited states which in a one-particle scheme lies in the range of ($E_F - |E_B/2| + \omega_0/2$, $E_F$), as it is shown in Fig. 15.

The dependence *E*(*k*) shown in Fig.15 corresponds to ARPES observations of kinks in a lot of HTSC materials (see, for example review [172]). For example, according to [172] in the well-studied cuprate Bi2212 the kink energy ($|E_B|/2$) is 70 meV.

The phonon nature of a kink is also pointed out by the observation of an isotopic effect in the vicinity of the kink energy [173], independence of the kink energy on the value of doping [174], independence of the kink energy on the nature of current carriers: according to [175], electron- and hole-doped cuprates have the same kink energy.

Fig. 16 shows a dependence of $dI/dV$ on *V* typical for HTSC which corresponds to the dependence of I on V presented in Fig. 15. There a kink corresponds to a dip on the curve to the right of the high peak.

Notice that since TI bipolarons exist for $T > T_c$ too, at temperature exceeding the critical one the $dI/dV$ curve will qualitatively retain the form shown in Fig. 16. Hence the quantity $|E_B|/2$ will play the role of a pseudogap in one-particle transitions, while $|E_B|$ – the role of a pseudogap in two-particle transitions. This conclusion is in full agreement with numerous tunnel experiments in HTSC [172],[176]-[177].

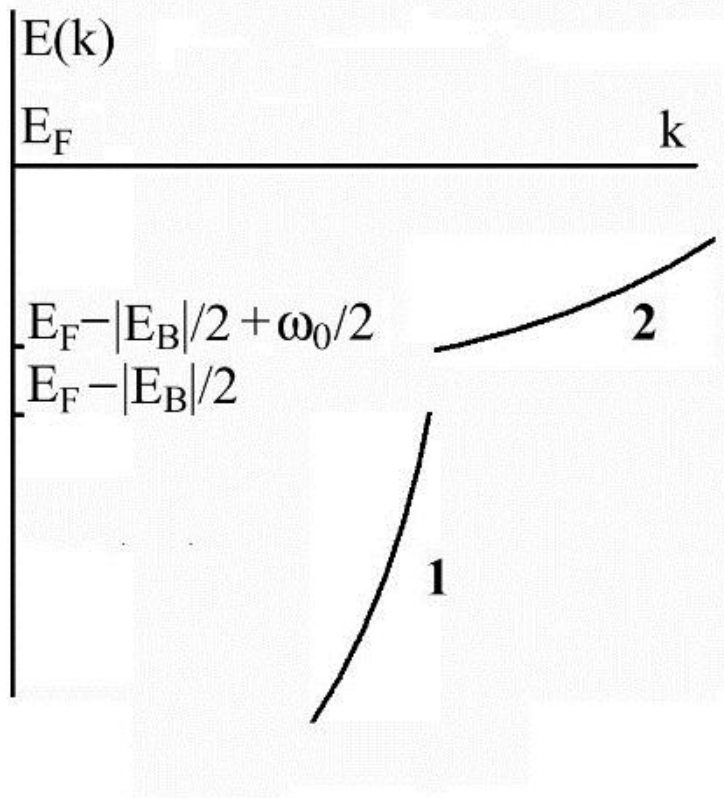

*Fig. 15.* Kink corresponds to a gap in passing on from normal branch 1 to TI bipolaron branch 2 for energy $E_F - |E_B/2|$

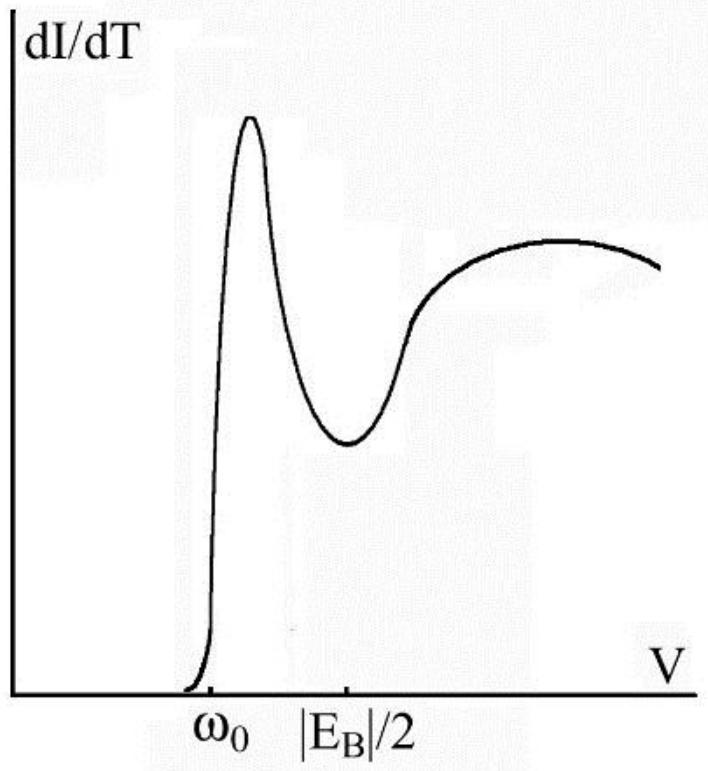

*Fig. 16.* Dependence of conductance $dI/dV$ on *V* corresponding to the volt-ampere characteristic shown in Fig. 14

**Angle Resolved Photo Electron Spectroscopy (ARPES).** Apart from STM, a direct method providing information on the properties of a superconducting gap is angle resolved photo electron spectroscopy [152]. Being added by the data of STM and the results of the quasiparticle interference, this method provides the most complete data on the properties of a SC gap. Recently, a method of double photoelectron spectroscopy has been developed where two electrons with certain momenta $\vec{k}_1$ and $\vec{k}_2$ and relevant energies $E_1$ and $E_2$ are emitted [178]. It is generalization of ARPES to the case of two particles. Despite the abundance of data obtained by ARPES the nature of a HTSC gap is still unclear. To a large extent this is due to the fact that up to the present time a unified theory of HTSC was lacking. If we proceed from the fact that a SC mechanism is caused by Cooper pairing then in the case of strong EPI this leads to the TI bipolaron theory if HTSC being considered. According to this theory, as distinct from bipolarons with broken symmetry, TI bipolarons are spatially delocalized and the polarization potential well is lacking (polarization charge is zero). According to Section 5, a TI bipolaron has a gap in the spectrum which has a phonon nature. In the TI bipolaron theory of SC, bipolarons are formed in the vicinity of the Fermi surface in the form of a charged Bose gas (immersed into electron gas) which condenses at the level lying lower the Fermi level by the value equal to the bipolaron ground state energy which leads to SC state. The spectrum of excitations of such a gas has a gap equal to the phonon frequency. In this section we will show that the photoemission spectrum obtained in ARPES just contains this gap and the gap $|E_B|/2$ determined from the two-particle current by STM which was considered in the previous section has nothing to do with the measurements of a gap by ARPES.

To this end we will proceed from the general expression for the light absorption intensity $I(\vec{k},\omega)$ measured in ARPES in the form:

$$I(\boldsymbol{k},\omega) = A(\boldsymbol{k},\omega)F(\omega)M(\boldsymbol{k},\omega). \tag{98}$$

In the case of the intensity of light absorption by TI bipolarons measured by ARPES, the quantities involved in (98) have a different meaning than in the case of a one-electron emission.

In the case of a Bose condensate considered, $\vec{k}$ has the meaning of a Boson momentum and $\omega$ is boson energy, $A(\vec{k},\omega)$ is a one-boson spectral function, $F(\omega)$ is a Bose-Einstein distribution function, $M(\vec{k},\omega)$ is a matrix element which describes transitions from the initial boson state to the final one.

In our case the role of a charged boson taking part in the light absorption belongs to a bipolaron whose energy spectrum is determined by (33), (40),[117],[118]:

$$\varepsilon_k = E_B\Delta_{k,0} + (E_B + \omega_0(\boldsymbol{k}) + k^2/2M)(1 - \Delta_{k,0}), \tag{99}$$

where $\Delta_{k,0} = 1$, if $k = 0$, $\Delta_{k,0} = 0$, if $k \neq 0$, whose distribution function is $F(\omega) = [\exp(\omega - \mu) - 1]^{-1}$. For $\vec{k} = 0$, a TI bipolaron is in the ground state, while for $\vec{k} \neq 0$ – in the excited state with energy $E_B + \omega_0(\vec{k}) + k^2/2M$, where $\omega_0(\vec{k})$ is a phonon frequence depending on the wave vector, $M = 2m$, $m$ is the electron effective mass.

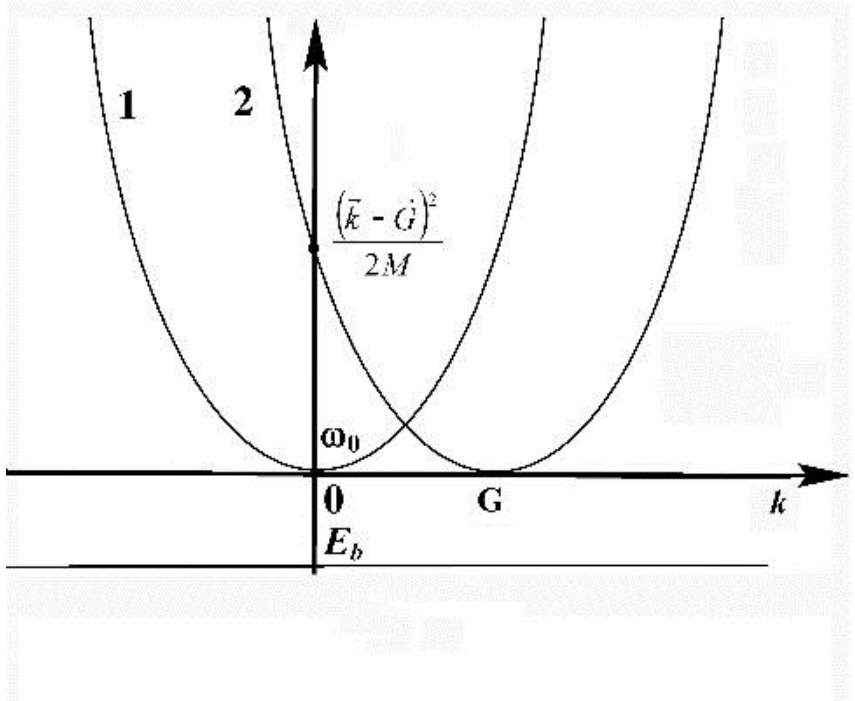

*Fig. 17* Schematic representation of the bipolaron transition to the excited state as a result of absorption of a light quanta

For further analysis it should be noted that the energy of bipolaron excited states reckoned from $E_B$ in equation (99) can be interpreted as the energy of a phonon $\omega_0(\vec{k})$ and the kinetic energy of two electrons coupled with this phonon. The latter, in the scheme of extended bands, has the form $(\vec{k}+\vec{G})^2/2M$, where $\vec{G}$ is the lattice inverse vector (Fig. 17). ARPES measures the spectrum of initial states which in our case is the spectrum of low-lying excitations of a TI bipolaron. In this connection we can neglect the contribution of one- and two-particle excitations of the electron (polaron) gas into which the bipolarons are immersed since the density of the TI bipolaron states in the vicinity of their ground state is much greater than that of the electron spectrum states. Hence, we *a priori* exclude consideration of such phenomena as the de Haas–van Alphen oscillation and the Shubnikov–de Haas oscillation [179]-[181]. Since the kinetic energy corresponding to the inverse lattice vector (or the whole number of the inverse lattice vectors) is very high, out of whole spectrum of a bipolaron determined by (99), we should take account only of the levels $E_B$ with $k=0$ and $E_B+\omega_0(\vec{k})$ with $k\neq 0$ as a spectrum of the initial states. In other words, with the use of the spectral function $A(\omega,\vec{k})=-(1/\pi)\mathrm{Im}G(\omega,\vec{k})$, where $G(\omega,\vec{k})=(\omega-\varepsilon_k-i\varepsilon)^{-1}$ is the Green bipolaron function, the expression for the intensity (98) can be presented as:

$$I(\boldsymbol{k},\omega)\sim\frac{1}{(\omega-E_B)^2+\varepsilon_1^2}\cdot\frac{1}{(\omega-E_B-\omega_0(\boldsymbol{k}))^2+\varepsilon_2^2}\ , \qquad (100)$$

which is fitting of the distribution function $F$ with $\mu=E_B$ and Green function $G$ by Lorentzians where $\varepsilon_1$ and $\varepsilon_2$ determine the width of the Bose distribution and bipolaron levels, respectively, (matrix element $M(\vec{k},\omega)$ involved in (98) has a smooth dependence on the energy and wave vector, therefore this dependence can be neglected).

Hence, as a result of light absorption by a pair of electrons (which are initially in a bipolaron state) ARPES measures the kinetic energy of electrons with momenta $k_e$, which are expelled from the sample in vacuum as a result of absorption of a photon with energy $\hbar\nu$. The energy conservation law in this case takes on the form:

$$\hbar\nu=\omega_0(\boldsymbol{k})+\frac{(\boldsymbol{k}+\boldsymbol{G})^2}{2M}=\xi+\frac{k_e^2}{m_0}+\omega_0(\boldsymbol{k}), \qquad (101)$$

$$\xi=2\Phi_0+|E_B|,$$

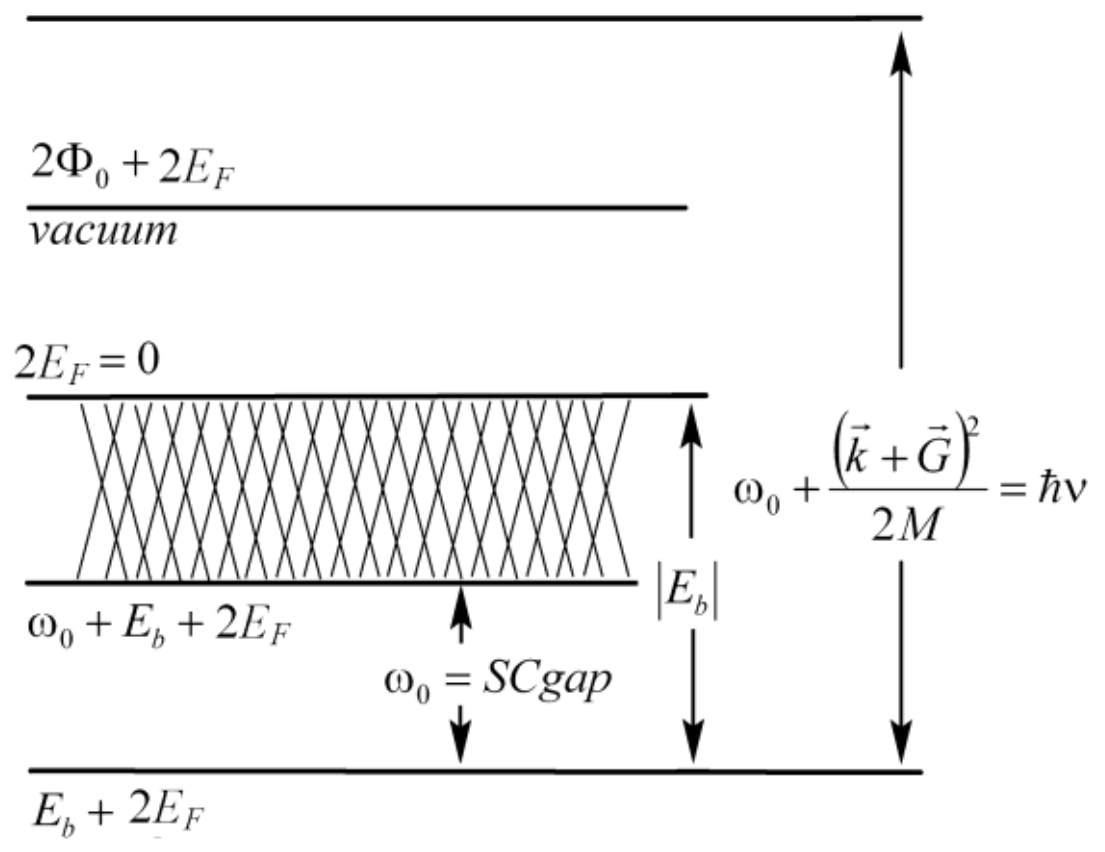

*Fig. 18* Scheme of energy levels in measuring the spectrum by ARPES. The region of the continuous spectrum lying below the Fermi level is shaded.

which is illustrated by Fig. 18, where $\Phi_0$ is the work of electrons escape from the sample, $m_0$ is the mass of a free electron in vacuum. Fig. 18 suggests that when a bipolaron is formed in the vicinity of the Fermi energy $E_F$ energy of two electrons becomes equal to $2E_F + E_B$.

In this case electrons pass on from the state with $p_F$, where $p_F$ is the Fermi momentum, to a certain state with momentum *p* below the Fermi surface (since $E_B < 0$). ARPES measures the spectrum of initial states reckoned from the energy $2 \cdot p^2 / 2m$ which corresponds to the energy of two electrons with momentum *p*. As a result, ARPES measures the energy $\omega_i = 2E_F + E_B - p^2 / m$.

Hence, if a bipolaron with energy $\omega = \omega_i = E_B + 2\vec{v}_F(\vec{p}_F - \vec{p})$, lying in domain of existence of a bipolaron gas $(2E_F + E_B, 2E_F)$, where $\vec{v}_F$ is the velocity of a Fermi electron, absorbs a photon with energy $\hbar\nu$, then a phonon arising as a result of the bipolaron decay is recorded in ARPES as a gap $\omega_0(\vec{k})$ and two electrons with the kinetic energy $k_e^2 / m_0$, determined by (101), are emitted from the sample.

In this scenario in each act of light absorption two electrons with similar momenta are emitted from the sample. This phenomenon can be detected by ARPES if the electron detector is placed just on the sample surface since the kinetic energy of flying of the emitted electron pair in vacuum (not compensated by the attracting potential in the bipolaron state) is several electron-volt.

Hence, ARPES, as discussed above, measures the phonon frequency $\omega_0(k)$ which is put into correspondence to the SC gap and therefore in cuprate HTSC with $d_{x^2-y^2}$ symmetry its angular dependence is determined by the expression $\omega_0(\vec{k}) = \Delta_0 |\cos k_x a - \cos k_y a|$.

From the viewpoint of phonon spectroscopy, identification of phonon modes of this type is difficult in view of their small number (equal to the number of bipolarons) as compared to the number of ordinary phonons equal to the number of atoms in a crystal. The spectral dependence of phonon frequencies is determined by both ion-ion interactions and an interaction with the electron subsystem of the crystal. Calculation of normal oscillations for a plane square lattice of atoms without taking account of the electron contribution leads to d-symmetry of their spectrum [128],[182]. With regard to $CuO_2$ SC planes of oxide ceramics, in the direction of Cu – O – Cu bonds (antinodal direction), phonons will have a gap, while in the direction of Cu – Cu bonds, i.e. along the unit cell diagonal (nodal direction) a gap will be lacking.

In calculating the electron contribution into the phonon spectrum account should also be taken of the relation between the electron density distribution and the position of ions on $CuO_2$ plane observed in STM/STS experiments with high spatial resolution [183].

The angular dependence $\omega_0(k)$ leads to the angular dependence of the intensity $I(\omega_i, \vec{p})\sim A(\omega_i, \vec{p})$ determined by equation (100) (Fig. 19) which is usually observed in ARPES experiments [152], [184, [185]. The form of the $I(\omega_i, \vec{p})$ dependence suggests that there is also a dependence of the absorption peaks on $\vec{p}$ symmetric about the Fermi level. This dependence is not presented in Fig. 19, since in view of a small population density of states with $p > p_F$ their absorption intensity will be very small [186].

Experimental checking of the effect of TI bipolaron emission as a whole is important for understanding the pairing mechanisms. Thus, according to [185], only one electron should escape from the sample with dispersion of the initial states determined, for $\vec{p} \neq 0$, by the formula $\varepsilon_p^{Bog} = \sqrt{(p^2/2m - E_F)^2 + \Delta^2(p)}$ (where $\varepsilon_p^{Bog}$ of a Bogolyubov quasiparticle), different from spectrum (99).

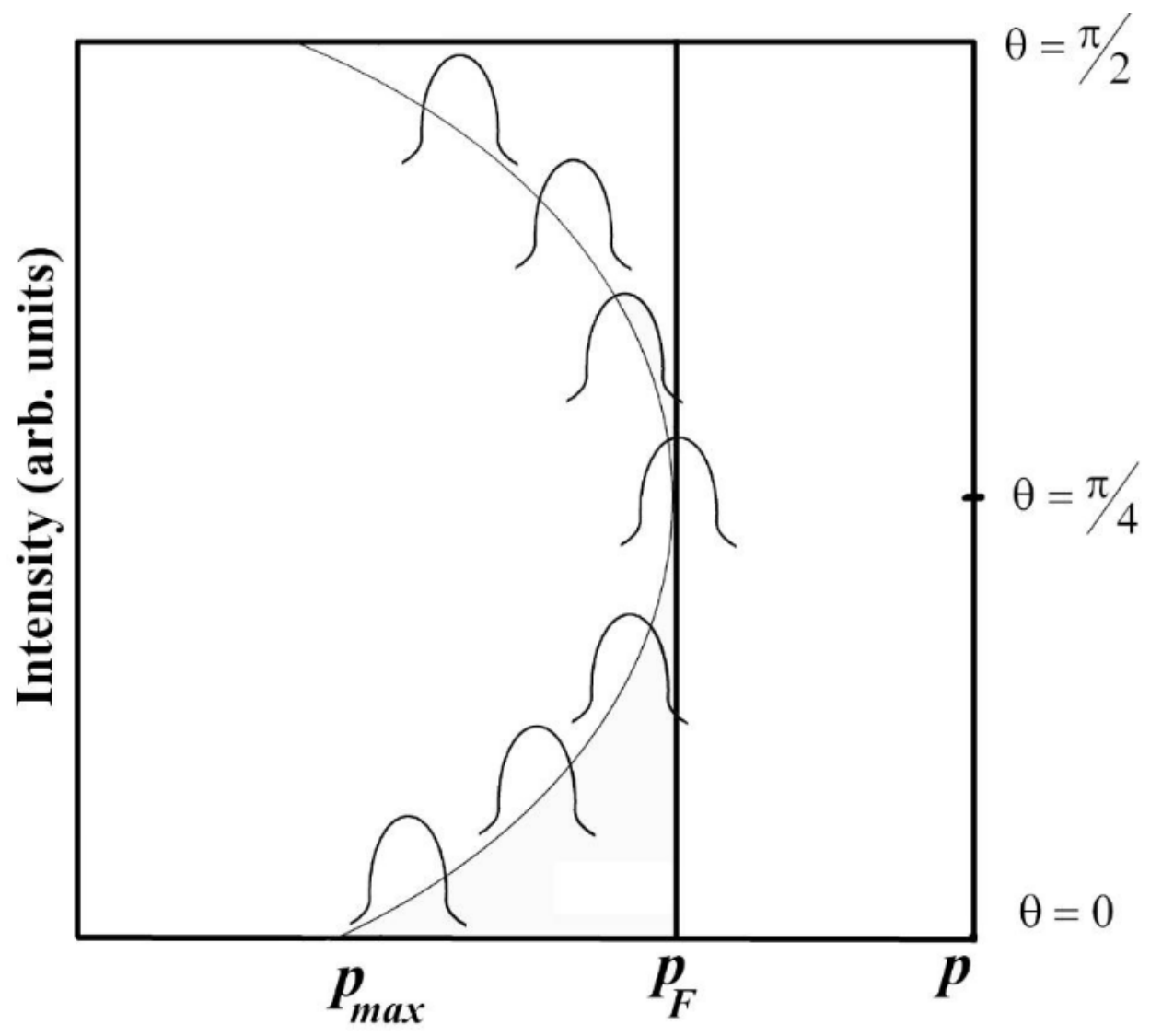

*Fig. 19.* Schematic representation of the angular dependence of the absorption intensity determined by (100) for $\omega = \omega_i$

The use of spectra $\varepsilon_P^{Bog}$ and (99) to describe the angular dependence of the intensity leads to a qualitative agreement with the ARPES data with currently accessible resolution. Experiments with higher resolution should give an answer to the question of whether a SC condensate in cuprates has fermion or TI bipolaron character.

The spectrum $\omega_0(k)$ suggests that in cuprate superconductors EPI constant becomes infinite in the nodal direction. Hence for bipolarons, a regime of strong coupling takes place in this case. Fig. 19 shows a typical dependence of the absorption intensity $I(\omega_i, \vec{p})$ observed in ARPES experiments [184].

The dependence shown in Fig.20 is obtained from the expression for the intensity (98) where the spectral function corresponds to the TI bipolaron spectrum (99) which cannot be obtained from spectral function (100) from paper [186] where Bogolyubov spectrum $\varepsilon_P^{Bog}$ is used for the spectrum and Fermi distribution function is used for the Bose distribution $F(\omega)$. This result can be considered as an argument in favor of a TI bipolaron mechanism of SC.

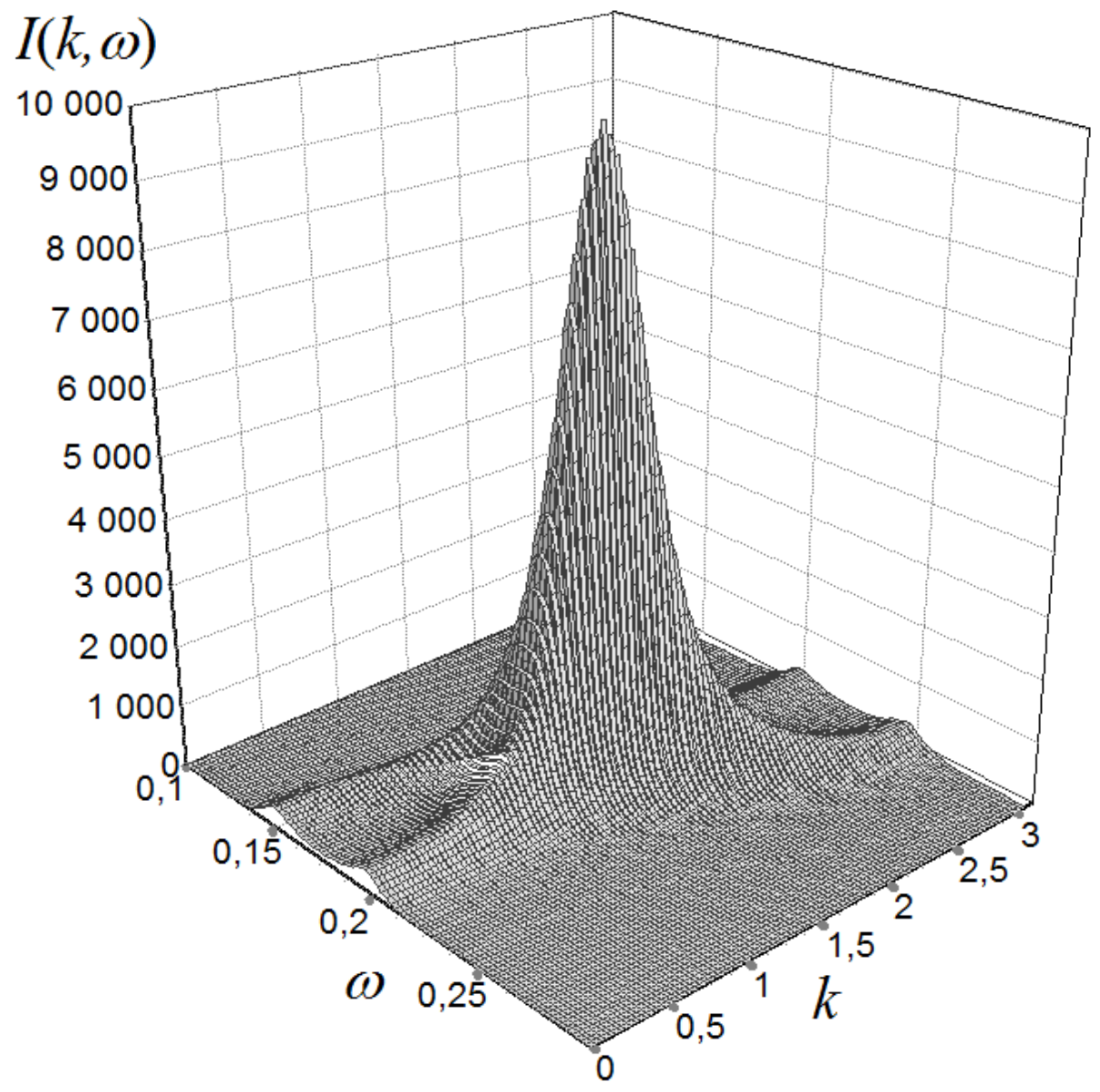

*Fig. 20.* Dependence of the absorption intensity $I(k,\omega)$ (arbitrary units) on $k$ and $\omega$ (eV), determined by (100), for the parameters: $|E_B| = 0{,}2\,\text{eV}$, $\Delta_0 = 0{,}05$ eV, $\varepsilon_1 = \varepsilon_2 = 0{,}01$ eV and the wave vector $k$ in antinodal direction. The lattice constant is assumed to be equal to one.

The peculiarities of the ARPES absorption spectrum considered above will also manifest themselves in tunnel experiments in the form of a thin structure (kinks) on the volt-ampere characteristics measured. To observe these peculiarities, as distinct from traditional ARPES measurements with high-energy photon sources ($\hbar\nu = 20 - 100$ eV), one should use low-energy photon sources ($\hbar\nu = 6 - 7$ eV) with higher momentum resolution [187]-[190].

In [191] a theoretical possibility to observe the emission of Cooper pairs by ARPES was considered for conventional SC. In particular, the authors of [191] demonstrated the availability of a peak in the emission current of Cooper pairs which corresponds to zero coupling energy of occupied two-electron states. The peak considered in [191] corresponds to a transition with energy $\hbar\nu$, determined by (101) where the coupling energy is ~1 meV, which is at the edge of ARPES accuracy. In the case of high-temperature superconductors the coupling energy can be ten times higher which makes checking of the effects considered more realistic. The main distinction of the results obtained here from those derived in [191] is the presence of the angular dependence of the absorption peak in Fig. 19,20, which is characteristic for HTSC materials.

Let us briefly discuss the temperature dependence of the intensity $I(\omega_i, \vec{p})$. According to equation (98), it is determined by the temperature dependence $F(\omega)$.

For $T < T_c$, where $T_c$ is the temperature of a SC transition $F(\omega) \cong N_0(T)$ for $\omega = E_B$, where $N_0(T)$ is the number of bosons (bipolarons) in a condensate which determines the temperature dependence of the absorption intensity. The value of $N_0(T)$ decreases as $T$ grows and, generally speaking, vanishes at the SC transition temperature making the absorption intensity vanish. Actually, however, this is not the case since only the Bose-condensate part vanishes. According to the TI bipolaron theory of SC, for $T > T_c$, bipolarons exist in the absence of a condensate too. In this case the population density of the ground state of such bipolarons will decrease as the temperature grows vanishing at $T^*$ which corresponds to a transition from the pseodogap state to the normal one.

This conclusion is confirmed by ARPES experiments in SC and pseudogap phases [192] which demonstrated that the angular dependence of a *d*-type SC gap is similar to the angular dependence of the state density in the pseudogap phase. At the same time there are considerable differences between the ARPES experimental data obtained for the SC phase and the gap one. In the SC phase the peak of absorption intensity occurs below the Fermi level which corresponds to a sharp spectral peak of the density of Bose-condensate states determined by equation (100), while in the pseudogap phase this peak is lacking in view of the lack of a condensate in it [153]. Under these conditions, since the population density of bipolaron excited states grows with growing temperature, the intensity of the absorption peak in ARPES experiments will decrease with growing temperature and reach minimum in the antinodal direction and maximum in the nodal one.

**Combination (Raman) scattering.** Though the combination scattering does not provide an angular resolution [193], its results also testify to the phonon nature of a gap in HTSC. As it was shown in [117], [118], the spectrum determined by (99), can be interpreted as a spectrum of renormalized phonons. Scattering of light with frequency $\nu$ on such phonons will lead to an appearance of satellite frequencies $\nu_+^B = \nu + \left|\varepsilon_k^B\right|$ and $\nu_-^B = \nu - \left|\varepsilon_k^B\right|$ in the scattered light, where $\varepsilon_k^B$ is determined by (99). In the case of wide conductivity bands, i.e. when the inequality $G^2/M >> \max\omega_0(k)$ is fulfilled, splitted lines $\nu_\pm^B$ overlap and form a region with a maximum displaced toward the Stocks branch $\nu_-$. Since in the model considered the bipolaron gas is placed into the polaron gas where the number of bipolarons is far less than the number of polarons, the intensity of bipolaron satellites will be much weaker than the intensity of TI polaron satellites: $\nu_+^P = \nu + \left|\varepsilon_k^P\right|$ and $\nu_-^P = \nu - \left|\varepsilon_k^P\right|$, $\varepsilon_k^P = E_P\Delta_{k,0} + \left(\omega_0 + k^2/2m\right)\cdot\left(1-\Delta_{k,0}\right)$, $E_P$ is the energy of a TI polaron. As in the case of usual combination scattering, the intensity of scattering on the polarons and bipolarons will be much weaker than the intensity of Rayleigh scattering corresponding to frequency $\nu$.

Indeed, in the combination scattering experiments [194], at $T < T_C$ a wide peak appears which, according to our interpretation, corresponds to widened frequencies $\nu_\pm^{B,P}$. In full agreement with the experiment, the position of this peak is independent of temperature. In the theory of the combination scattering based on the BCS, on the contrary, the position of the peak should correspond to the width of the SC gap and for $T = T_C$ the frequency corresponding to this width should vanish.

The combination scattering results also confirm that TI bipolarons do not decay at $T = T_C$, but persist in the pseudogap phase. Measurement of the temperature dependence of the combination scattering intensity is based on the subtraction of the absorption intensity in the normal and superconducting phases. The difference obtained, according to our approach, is fully determined by scattering on the Bose condensate and depends on temperature vanishing at $T = T_C$.

## 13. Conclusive remarks

In the theory presented, as in the BCS, the momentum of the bipolaron mass center $\vec{\mathcal{P}}$ (accordingly $\vec{G}$ in a magnetic field) is equal to zero. According to this theory, the SC state is a homogeneous bipolaron BEC. The theory can be generalized to the case of a moving BEC with $\mathcal{P}\neq 0$, which remains homogeneous when moving. In this case some interesting peculiarities arise [195]. Presently, a wide discussion is devoted to the possibility of the formation of a inhomogeneous BEC in the form of the so called pair density waves (PDW) which destroy the translation invariance [154],[196]-[200]. However, here the situation is different from the problem of polarons and bipolarons with broken or actual TI symmetry (Sections 1, 2). The scenario of SC with PDW including the presence of charged density waves (CDW) or spin density waves (SDW) [201]-[204] is provided by the discreteness of the crystal which is not taken into account in the continuum model of EPI. The problem of the competition between the CDW mechanism of SC and the bipolaron one is considered, for example, in [205] for a SRP in squeezed vacuum.

Modulation of BEC density for wave vector corresponding to nesting leads to the appearance of a gap in the spectrum which in many works is identified with a SC gap [154],[196]-[204]. In this case the TI bipolaron gap $\omega_k$, being universal, would have the properties of a pseudogap manifesting itself as a low-energy thin structure in the conductance spectrum of optimally doped SC [166]-[168].

In the approach considered we actually did not use any specificity of the mechanism of the electron or hole pairing. For example, both in the Hubbard model and in the $t-J$ model, in describing copper oxide HTSC the same holes take part in the formation of antiferromagnetic fluctuations and pairing caused by an exchange by these fluctuations. If an interaction of holes with magnetic fluctuations leads to the formation of TI magnetopolarons having the spectrum $\omega_0(k)$, then this spectrum is also the spectrum of magnons renormalized by their interaction with holes (bound magnons). For this reason the statement that the RVB superconductor is just a limiting case of the BCS SC with strong interaction becomes justified [206] (in this case the role of polarons and bipolarons belongs to holons and biholons).

Evidently, *d*-symmetry is specificity of cuprate HTSC and is not a precondition of the existence of HTSC. For example, sulphide ($H_3S$) demonstrates a record value of the transition temperature: $T_C = 203\,\mathrm{K}$ (under high pressure [207]), does not have a magnetic order, but EPI is strong in it. Still greater value of $T_C$ under high pressure has recently been obtained in the substance $LaH_{10}$ with $T_C = 260\,\mathrm{K}$ [208], where the EPI is also strong and a magnetic order is lacking.

Nevertheless, the mechanism of pairing is still unclear. If it is provided by an interaction of current carriers with magnetic fluctuations, then, in the approach considered, the particles which bind electrons into pairs will be magnons rather than phonons. In passing on from the pseudophase to the normal one this binding mode disappears which leads to the decay of a bipolaron into two individual polarons with the emission of a phonon (magnon).

In the pseudogap phase there may be a lot of different gaps caused by the presence of phonons, magnons, plasmons and other types of elementary excitations. In this case the SC gap will be determined by the type of elementary excitations whose interaction with the current carriers is the strongest.

From the viewpoint of the TI bipolaron theory a possible resultant picture of HTSC looks as follows.

According to the above consideration, the foundation of the microscopic theory is provided by the TI bipolaron EPI mechanism. It follows from the theory that in order to reach high $T_c$ one should primarily enhance the concentration of TI bipolarons. In oxide ceramics this is reached by the presence of antiferromagnetic order and stripes in them.

Playing the role of microscopic domain walls, the stripes, having a ferromagnetic order, attract electrons. Due to exchange interaction the energy of electrons in the stripes is lower than that in the rest of the template (analog of ferrons by Nagaev [60] with regard to the contribution of polaron [209] and magnetostriction effects [210] into their formation), accordingly, the concentration of electrons there is rather high. To restore a charge equilibrium TI bipolarons flow from the stripe regions to the template thus enhancing the concentration of TI bipolarons in it and, on the whole, $T_c$ of the sample. This redistribution gives rise to a PDW (elevated concentration of bipolarons in the template and reduced concentration in the stripes) and CDW (elevated concentration of electrons in the stripes and reduced concentration in the template).

The mechanism described enables one to construct purposively SC materials which could work at room $T_c$. As it was pointed out in [118], to do so one can use inhomogeneous doping making the periphery of a HTSC cable doped with ferromagnetic impurities which could attract electrons from the core of the cable. As a result one can take a cable with enhanced concentration of TI bipolarons on its axis and, as a consequence, high $T_c$.

The theory of SC on the basis of EPI presented should rather be considered as a scheme for describing the properties of real materials. The number of different substances possessing HTSC properties is many thousands and the number of publications on HTSC – many hundreds of thousands. Therefore the construction of a microscopic theory of HTSC shall probably be understood to mean a certain ideological concept whose role can be played by the TI bipolaron mechanism considered in the review.

**14. Supplement**

Hamiltonian $H_1$ involved in (26) has the form:

$$\widehat{H}_1 = \sum_k (V_k + f_k \hbar\omega_k)\,(a_k + a_k^+) + \sum_{k,k'} \frac{\boldsymbol{kk'}}{m} f_{k'} \left(a_k^+ a_k a_{k'} + a_k^+ a_{k'}^+ a_k\right) + \frac{1}{2m} \sum_{k,k'} \boldsymbol{kk'}\, a_k^+ a_{k'}^+ a_k a_{k'}.$$

Let us act on the functional $\hat{\Lambda}_0|0>$, by the operator $\widehat{H}_1$ where $\hat{\Lambda}_0$ is an operator which generates Bogolyubov-Tyablikov canonical transformation (29). Let us show that $< 0|\Lambda_0^+ H_1 \Lambda_0|0 >= 0$. Indeed, the action of $\Lambda_0$ on the terms of $H_1$, which contains an odd number of operators $a_k, a_k^+$ (the first and second terms in $\widehat{H}_1$) yields an expression which contains an odd number of these operators and the expectation for this expression well be equal to zero.

Let us consider expectation for the last term of $H_1$:

$$< 0|\hat{\Lambda}_0^+ \sum_{k,k'} \boldsymbol{kk'} a_k^+ a_{k'}^+ a_k a_{k'} \hat{\Lambda}_0|0 >. \qquad \text{(S1)}$$

The functional $< 0|\hat{\Lambda}_0^+ a_k^+ a_{k'}^+ a_k a_{k'} \hat{\Lambda}_0|0 >$ is the norm of the vector $a_k a_{k'} \hat{\Lambda}_0|0 >$ and therefore is positively determined for all $\vec{k}$ and $\vec{k}'$. If we replace $\vec{k} \to -\vec{k}$, the whole of the expression will change the sign. Therefore (S1) is equal to zero.

As it was shown by Tulub [75], [96] the operator $\hat{\Lambda}_0$ is generator of Bogolyubov-Tyablikov transformations and has the form:

$$\hat{\Lambda}_0 = c \exp\left\{\frac{1}{2} \sum_{k,k'} a_k^+ A_{kk'} a_{k'}^+\right\},$$

where *c* is a normalizing constant and the symmetric matrix A satisfies the conditions:

$$A = M_2^* (M_1^*)^{-1}, \;\; A = A^T,$$

where $M_1$ and $M_2$ are Meller matrices [211] involved in Bogolyubov-Tyablikov transformation (29).